\documentclass[prd,aps,showpacs,amsmath,amssymb,twocolumn,nofootinbib]{revtex4-1}
\usepackage{graphicx}
\usepackage{dcolumn}
\usepackage{bm}
\newcommand{\Fstar}{\raisebox{.2ex}{$\stackrel{*}{F}$}{}}

\begin{document}

\title{Probing nonlinear electrodynamics in slowly rotating spacetimes through neutrino astrophysics}

\author{Herman J. Mosquera Cuesta $\mbox{}^{1,2}$}
 \email{herman@icra.it}
 \author{Gaetano Lambiase $\mbox{}^{3,4}$}
 \email{lambiase@sa.infn.it}
 \author{Jonas P. Pereira$\mbox{}^{5}$}
 \email{jonas.pereira@ufabc.edu.br}
\affiliation{$\mbox{}^{1}$Visiting Scientist of Colciencias, Programa Nacional de Ciencias B\'asicas/Ciencia Espacial \\ Av. Calle 26 No. 57 - 41 Torre 8, Bogot\'a, Colombia} 

\affiliation{$\mbox{}^{2}$ Instituto Federal de Educa\c c\~ao, Ci\^encia e Tecnologia
do  Cear\'a, Avenida Treze de Maio, 2081, Benfica, Fortaleza/CE, CEP 60040-531, Brazil}
\affiliation{$\mbox{}^{3}$Dipartamento di Fisica ``E. R. Caianiello'', Universit\`{a} di Salerno, 84084 Fisciano (SA), Italy}
\affiliation{$\mbox{}^{4}$INFN--Gruppo Collegato di Salerno, Italy}
\affiliation{$^{5}$Centro de Ci\^encias Naturais e Humanas, Universidade Federal do ABC, Avenida dos Estados 5001, CEP 09210-580, Santo Andr\'e, SP, Brazil}
\date{\today}

\begin{abstract}
Huge electromagnetic fields are known to be present during the late stages of the dynamics of supernovae. Thus, when dealing with electrodynamics in this context, the possibility may arise to probe nonlinear theories (generalizations of the Maxwellian electromagnetism). We firstly solve Einstein field equations minimally coupled to an arbitrary (current-free) nonlinear Lagrangian of electrodynamics (NLED) in the slow rotation regime $a\ll M$ (black hole's mass), up to first order in $a/M$. We then make use of the robust and self-contained Born-Infeld Lagrangian in order to compare and contrast the physical properties of such NLED spacetime with its Maxwellian counterpart (a slowly rotating Kerr-Newman spacetime), especially focusing on the astrophysics of both neutrino flavor oscillations ($\nu_e \rightarrow \nu_\mu, \nu_\tau$) and spin-flip ($\nu_l \rightarrow \nu_r$, ``$l$'' stands for ``left'' and ``$r$'' stands for ``right'', change of neutrino handedness) mass level-crossings, the equivalent to gyroscopic precessions. Such analysis proves that in the spacetime of a slowly rotating nonlinear charged black hole (RNCBH), intrinsically associated with the assumption the electromagnetism is nonlinear, the neutrino dynamics in core-collapse supernovae could be significantly changed. In such astrophysical environment a positive enhancement (reduction of the electron fraction $Y_e<0.5$) of the r-process may take place. Consequently, it might result in hyperluminous supernova explosions due to enlargement, in atomic number and amount, of the decaying nuclides. Finally, we envisage some physical scenarios that may lead to short-lived charged black holes with high charge-to-mass ratios (associated with unstable highly magnetized neutron stars) and ways to possibly disentangle theories of the electromagnetism from other black holes observables (by means of light polarization measurements).
\end{abstract}

\maketitle

\section{Introduction}
It is well-known that particles endowed with spin also interact with gravity \cite{1972gcpa.book.....W}. In the astrophysical scenario, the most commonly observed ones are photons, though neutrinos are also produced bountifully in coalescing systems (see \cite{Nus-from-binary-NSs-Mergers} and references therein) due to nuclear fusion reactions in the nucleosynthesis of heavy elements \cite{1996NuPhA.606..167F}, playing a very important role in the supernova physics. Neutrinos subsist just on superposition of mass eigenstates: flavor states \cite{1978PhR....41..225B}. This aspect is noteworthy since it implies flavors can oscillate under convenient conditions in physical (labs, accelerators) and astrophysical environments (exploding stars),  which may lead to observable effects \cite{long-list-of-SN-Nu-Physics-Articles}. This aspect was exactly the early reason for the introduction of the flavor states, in order to lead to the neutrino oscillations that could explain the theretofore anomalous abundance of neutrinos coming from the Sun \cite{2004lpih.conf..385B, 2012PhRvL.108e1302B}, as well as the ones present in material media \cite{M-Smirnov, Wolfenstein1978-1979}. Due to the improvements in detecting neutrinos, e.g. MiniBooNE at Fermilab USA, KamLAND-Zen collaboration, CERN/Geneve - Gran Sasso/Italy, ANTARES in the France  Mediterranean sea, SuperKamiokande and K2K in Japan, Baksan neutrino observatory (BNO) in Caucasus mountains in Russia, Daya Bay reactor neutrino experiment in China, Sudbury neutrino observatory (SNO) in Ontario Canada, IceCube neutrino observatory at the South Pole \cite{Abbasi2009-2011}, etc., detailed analyses where neutrino conversions take place become more pertinent. More importantly yet, due to the unavoidable interaction of neutrinos with gravity, such particles could give us invaluable and precise information about various astrophysical environments such as exploding stars, i. e. supernovae.

In a supernova, neutrinos carry away almost all the binding energy of the just-born neutron star,  i.e.,  $\Delta  E_\nu  \sim  3\times  10^{53}$~erg
\cite{G-Mathews-Review-2014}. Because of this abrupt neutrino cooling process (which may lead the system to increase its density) the proto-neutron star (PNS) might undergo a catastrophic phase transition to hybrid or quark star, where an interacting strange kaon condensation state is said to appear \cite{Catastropic-PT-Greiner-etal} \footnote{The reason for this lies mainly in the fact that densities so high (supranuclear) could be attained during the gravitational collapse that even a quark phase might arise \cite{Glendenning-book}, as suggested by QCD physics involving the appearance of Cooper pairs,  ``bosonization'' and/or kaon condensation, color-flavor locking, quark deconfinement, and other theoretically allowed QCD stages which might drive phase transitions at the very inner core (for further details see \cite{Haensel-book} and references therein). These stages could appear when a equation of state of cold baryon-rich matter is considered. All this would be as well accompanied by the expected neutrino cooling which would decrease even more the pressure of the supernova progenitor (deleptonization process), allowing this way for gravity to make the star even further compact.}, possibly leading
to the formation of a short-lived rotating and nonlinear charged black hole (RNCBH) \cite{HJMC-NLED-CBH-Formation, Brown-Bethe-1994}, that is, a black hole described by a theory of the electromagnetism more general than Maxwell's. This might happen because electromagnetic fields could easily surpass certain scale fields (intrinsically associated with nonlinear theories of the electromagnetism) near the black hole horizons if they are charged enough. In Sec. \ref{summary} we envisage some scenarios which might result in such situation. The RNCBH may appear, for instance, after the just-formed proto-neutron star undergoes a phase transition creating a charge separation amidst the crust and the collapsing core \cite{HJMC-NLED-CBH-Formation}, as well as by means other possible effects, which will also be briefly discussed in Sec. \ref{summary}. 

All the above motivates and leads us to the  main scope of this work: to surmise the existence of axially symmetric nonlinear charged black holes (at least for some instants of time, i.e. transiently) and study their properties and implications, specially through the physics of neutrinos. Foremost, we solve the system of equations coming from the minimal coupling of standard general relativity with generalizations of the Maxwell's Lagrangian, known as nonlinear electrodynamics (NLED), in the slow rotation regime ($a\ll M$, $a$ the rotational parameter of the black hole and $M$ its mass). 

NLED is the approach to describe electromagnetic interactions in a relativistically invariant set up. Several approaches were envisioned: Heisenberg; Euler and Kochel; Euler; Heisenberg and Euler (added $F^2$-term) \cite{1935NW.....23..246E, 1936AnP...418..398E, 1936ZPhy...98..714H, 2012IJMPS..14...42D, 2013RPPh...76a6401B}; and Weisskopf (added a logarithmic-term) \cite{Weisskopf1936}; Born; Born and Infeld \cite{1934RSPSA.144..425B} (bounded the electric field strength by giving to the electron a finite radius), Plebanski (robust framework, including plasma physics) \cite{plebanski}, to extend Maxwell electrodynamics (linear in Lorentz invariant $F$) so as to deal with divergences in analysis of electromagnetic (EM) phenomena, {as well as to insert desired effects into the theory under the classical point of view}. Applications of NLED have been extensively studied in the literature,
extending from cosmological and astrophysical contexts \cite{2011JCAP...03..033M, 2011APh....34..587C, 2009PhRvD..80b3013M, 2004MNRAS.354L..55M, 2004ApJ...608..925M, 2006IJMPA..21...43C, 2007EL.....7719001M, 2008MNRAS.389..199M}, to nonlinear optics \cite{2006hep.th...10088D}, high power laser technology and plasma physics \cite{2006RvMP...78..591M, 2006PhPl...13j2102L, 2006PhRvL..96h3602L, 2012IJMPA..2760010H}, and the field nonlinear exponential growth due to chiral
plasma instability during the weak parity-violating electron-capture (chirality imbalance) process in core collapse SNe \cite{2014arXiv1402.4760O, 2013PhRvL.111e2002A}. These authors stress that the original B-field gives a positive feedback to itself, to grow exponentially, being this last the actual chiral plasma instability. In our understanding, this field increase would suggest that NLED might be at action inside just-born pulsars. For further details on magnetic field amplification, see Ref. \cite{M.Shibata-etal-2015}. 

Meanwhile, the gravitational effects on the neutrino oscillation phases (between active species $\nu_a \longrightarrow \nu_b$ and of active into sterile species $\nu_a \longrightarrow \nu_s$), and consequently on the overall neutrino dynamics (which would also include neutrino spin-flip conversions $\nu_l\rightarrow \nu_r$, related to their handedness), have been on focus of several discussions in the literature \cite{1996GReGr..28.1161A, 1997PhRvD..55.7960C, long-list-of-articles-on-Nu-Grav_Effects, 1996NuPhA.606..167F}. It has become clear that gravity would be
essential for building a complete picture of the neutrino dynamics in very dense and self-gravitating matter, in particular in the very deep inside regions of supernovae. In recent analyses \cite{long-list-of-articles-on-Nu-Grav_Effects, 1996NuPhA.606..167F}, it has been
pointed out that the neutrino outflow and related supernova expansion have been discussed
in most of the specific literature ignoring any gravitationally-induced effects. Indeed,
neutrino oscillations in the accretion disk produced by the coalescence
of a binary neutron star system \cite{Nus-from-binary-NSs-Mergers}, and in the inner edge of the fall-back supernova
ejecta, as well as
around the neutrinosphere, could be strongly influenced by gravity, which then would affect the supernova accretion dynamics (neutron digging in fingers) \footnote{For non experts in the field of numerical modeling of supernovae, in what follows we resume the relevant physics and astrophysics pertinent to the concept of \textsl{neutron fingers} taking benefit from the abstract of Ref. \cite{Bruenn-etal-2004}, and the discussion on this fluid feature given in Ref. \cite{Mackenzie-Warren-etal-2014}.
Neutron fingers are instabilities in a Ledoux stable fluid driven by thermal 
and lepton diffusion, technically quoted as \textit{doubly diffusive instabilities}.   
Whenever these fluid motions are present below the neutrino sphere in a  
core-collapse supernova progenitor, they can induce convective-like fluid 
motions at those supernova layers, and may enhance the neutrino emission 
by advecting neutrinos outward toward the neutrino sphere, what may thus 
play an important role in the supernova mechanism. Neutron fingers have 
also been suggested as being critical for producing explosions in the sophisticated spherically symmetric supernova simulations by the Livermore group. Such instability has been argued to arise in an extensive region below the neutrino sphere of 
a proto-supernova where entropy and lepton gradients are stabilizing and destabilizing, respectively, if, as that group asserts, the rate of neutrino-mediated thermal equilibration greatly exceeds that of neutrino-mediated lepton equilibration. According to Bruenn and collaborators \cite{Bruenn-etal-2004}, application of the Livermore 
group’s criteria to models derived from core collapse simulations using both their equation of state and the very well-known Lattimer-Swesty equation of state do show a large region below the neutrinosphere unstable to neutron fingers. Indeed, from the convective regions below the neutrinosphere, neutron fingers dig into the star and reach its center in about one second. Then they propagate outward to englobe almost all the exploding star. An interesting discussion on the relevance of this astrophysical fluid dynamics phenomenon and its timescale (1-50 ms) for the production of bursts of gravitational waves during the deleptonization phase of supernovae is given in Ref. \cite{HJMC-etal-2004}. } \cite{Bruenn-etal-2004, Mackenzie-Warren-etal-2014} and the
final explosion wind. Because of this, all the fundamental quantities of relevance for
the explosion dynamics would be in principle affected by the gravitational field
of the putatively just formed RNCBH here under analysis. Thus, most of the gravitational effects we
shall discuss in the present paper would be directly connected with the drawbacks or
difficulties of current neutrino oscillation description of the effects in supernova
explosions, which still lead to fail in succeeding to eject the stellar inner mantle,
especially in 3-D simulations \cite{Bruenn-etal-2004}. Regarding this issue SN modelers play to argue that most
likely turbulence is the culprit.

We therefore state hereafter that the full consideration of gravitational effects on
neutrinos propagating in the nearby spacetime of a black hole should be taken into account in SNe studies, and that a possibly relevant piece for engineering such process would be a rotating and charged black hole or any very compact object permeated by nonlinear electromagnetic fields.
For instance, the gravitational redshift (sensitive to nonlinear electrodynamics) should play a fundamental role in the
entire supernova explosion physics. And as both theory and observations indicate such properties should take place in the final stage of evolution of massive stars, e.g. a Wolf-Rayet, or the coalescence a binary NS system.

Summarizing, the central engine here purported should be properly integrated in any
scheme intended to successfully explain the dynamics of such astrophysical explosions.
This is the principal motivation of the present paper.

\subsection{A brief account on neutrino oscillations in a gravitational field}
As already mentioned, most of the dynamical features associated with neutrino flavor transformations are intimately connected to or dependent on their difference in masses $\Delta m^2_{21} \equiv m^2_2 - m^2_1$, or simply $\Delta m^2$. Therefore, in order for a flavor conversion to be an observable while happening over a distance $x$, in a curved spacetime, the wave packet describing a couple of mass eigenstates should overlap (i.e. they should undergo quantum-mechanical interference), otherwise each of the
individual masses will separate from each other as time goes by. The comments in this paragraph apply to the \textit{vacuum oscillations}, although it is clear that similar conditions apply to the level-crossing phenomenon of oscillations in matter, or MSW effect \cite{M-Smirnov, Wolfenstein1978-1979}. 
In this last case the difference of the squares of the neutrino mass eigenvalues ${\Delta}_{|msw} m^2$, the mixing angle in matter $\tan2 {\theta}_{|msw}$, and the resonance condition $v_{|msw}(r)$ should also be affected by the gravity associated with the curved spacetime.

In general relativity there exists a condition on the \textit{width} of the neutrino wave packets such that neutrino oscillations are observed while taking place in a curved spacetime. Recalling that the infinitesimal line element in such spacetime reads $ds^2 = g_{\alpha\beta} dx^\alpha dx^\beta$ (for simplicity we assume here a sperically symmetric spacetime in a coordinate system such that $g_{\alpha\beta}$ is diagonal and we choose the metric signature $[+,-,-,-]$), the searched condition on the \textit{width} of the neutrino wave packets $\Delta d$ translates into the covariant inequality \cite{1997PhRvD..55.7960C}
\begin{equation}
\Delta d \gtrsim \int (-g_{ij} P^i_2 P^j_2)^{\frac{1}{2}}  d\lambda - \int
(-g_{ij} P^i_1 P^j_1)^{\frac{1}{2}}  d\lambda,
\label{cond-nu-wave-packets}
\end{equation}
where $\lambda$ is an affine parameter along the geodesics and $P_i$, $i=1,2,3$, are the space components of $P_{\mu}$ (the conjugate four-momentum to $x^{\mu}$, the generator of spacetime translations of neutrinos), which satisfies the ``mass shell condition'' $P^{\mu}P_{\mu}=m^2$. As seems reasonable due to the neutrino very small masses, we approximate the tangent four-vectors to the trajectories, $\dot{x}^{\mu}\doteq d x^{\mu}/d\lambda$, to null-like ones, $\dot{x}^{\mu}\dot{x}_{\mu}=0$. Due to the freedom in re-scaling the affine parameter, we assume here that $P^0=\dot{x}^0$ and $P^i_a= \dot{x}^i(1 - \epsilon_a)$, $a=1,2$, with $\epsilon \ll 1$ due to the neutrino small masses and $a$ stands for the neutrino mass eigenstates. From $\dot{x}^{\mu}\dot{x}_{\mu}=0$, small $\epsilon$ and the mass shell relation, it straightforwardly follows that
\begin{equation}
\epsilon_a= \frac{m^2_a}{-2g_{ij} \dot{x}^i\dot{x}^j}=\frac{m^2_a}{2g_{00}(\dot{x}^0)^2}\label{epsilon_a}\,.
\end{equation}
Finally, from Eq. (\ref{epsilon_a}) and by recalling that $dl^2 = -g_{ij} \dot{x}^i\dot{x}^j (d \lambda)^2= g_{00} (\dot{x}^0)^2 (d \lambda)^2$, where $d l$ is the infinitesimal proper spatial length (for $t$ constant) in the spacetime $g_{\mu\nu}$, we have that Eq. (\ref{cond-nu-wave-packets}) can be cast as
\begin{equation}	 
\Delta d \gtrsim  \frac{\Delta m^2}{2} \int \frac{dl}{[g_{00}(\dot{x}^0)^2]}=\frac{\Delta m^2}{2} \int \frac{g_{00}}{E_0^2} dl,
\label{curved-ST-cond-nu-wave-packets}
\end{equation}
where $E_0$ is a constant (energy at infinity) coming from the geodesic equation $g_{00} \dot{x}^0=E_0$. 

As one can understand from the above brief analysis, to properly discuss the potential detectability of neutrino flavor conversions taking place over the spacetime of a nonlinear charged and slowly rotating black hole, one of the the goals of this paper, it is needed to having computed the $g_{00}$ metric component of such geometry, once the neutrino energy is known in advance. 
Consequently, because in our study case (neutrino physics inside the cores of supernovae) this geometry could be far different from the one corresponding to the Schwarzschild spacetime, one would expect to find not previously reported effects. This way, all the information that could be gathered in connection to such events may help characterize whether the supernova event formed a Schwarzschild-like black hole rather than a Kerr-Newman-like one, which would prove the astrophysical formation and existence of such compact supernova remnants. An idea relatively similar to this our view here involving neutrino propagation inside supernovae was also discussed by Beacom as the signature of the formation of a BH inside a SN \cite{Beacom-BH-SN}.

Our analyses in this work could also be seen under the following perspective: finding astrophysical entities and environments that could be used as tools to probe (nonlinear) electrodynamical processes in the cosmos, quite similarly, conceptually speaking, to Crispino and collaborators' investigations, who used scattered electromagnetic radiation to probe the charge of a black hole \cite{2014PhRvD..90f4027C}.

The plan of this paper is the following. In the next section we obtain the field equations for slowly rotating nonlinear charged black holes. Section \ref{geodesic} reviews the geodesics in the aforesaid spacetimes, important for the neutrino physics, such as flavor oscillations, that shall be investigated in Sec. \ref{flavoroscillation}, and spin precession (or spin-flip), discussed in Sec. \ref{spinprec}. In Sec. \ref{nosfbi} we apply the generic results of neutrino oscillations and spin precession for the Born-Infeld theory, in order to explore their differences when compared to the Maxwell Lagrangian. We also make use of the effect of frame dragging in axially symmetric spacetimes to contrast the aforementioned theories, done in Sec. \ref{nonlinprec}. Section \ref{rproc} is devoted to elaborate upon the relevance of charge (nonlinearity of the electromagnetism) in black holes for r-processes. Simple estimates are given in Sec. \ref{estimation} only for assessing relevant scales for some physical processes in the astrophysical context. Finally, in Sec. \ref{summary} we discuss and summarize the main points of our assumptions and analysis. We work with geometric units unless otherwise stated. For the electromagnetism we work with Gaussian units. The metric signature is chosen to be $-2$ [($+,-,-,-$)].

\section{Field equations for slowly rotating nonlinear black holes}
\label{feslow}
When one considers that the norm of the angular momentum per unit of mass, $a$, of a spinning black hole is constrained to be much smaller than its outer horizon $r_+$ {(which implies $a/r\ll 1$, as well as $a/M\ll1$, $M$ its mass)}, then, based on the Kerr-Newman solution \cite{1973grav.book.....M}, the Ansatz to the metric to account for nonlinear Lagrangians of the electromagnetism can be written in Schwarzschild coordinates $({t,r,\theta,\phi})$ as
\begin{eqnarray}
ds^2&=& g_{00}(r)dt^2-\frac{1}{g_{00}(r)}dr^2 -r^2d\theta^2- r^2\sin^2\theta
d\phi^2\nonumber\\ &-&2a\sin^2 \theta A(r)dt\, d\phi ,
\label{metric2}
\end{eqnarray}
where $g_{00}(r)$ is the solution to the associated static and spherically symmetric
black hole for the theory under interest. In Eq. (\ref{metric2}), $A(r)$ is function to be determined from the nonlinear electromagnetic field equations, which we describe below.

The whole set of field equations is obtained by the minimal coupling between standard general relativity (Einstein-Hilbert action) and nonlinear theories of the electromagnetism with Lagrangian densities dependent upon its invariants $L(F,G)$, and it reads (see \citep{2014PhLB..734..396P} and references therein) 
\begin{equation}
G_{\mu\nu}= 8\pi T_{\mu\nu},\;
\frac{\partial}{\partial x^{\mu}}[\sqrt{-g}(L_{F}F^{\mu \nu} + L_G \Fstar^{\mu\nu})]
= 0,
\label{d2}
\end{equation}
added to
\begin{equation}
\frac{\partial}{\partial x^{\mu}}
(\sqrt{-g} \Fstar^{\mu\nu})=0 \label{emid}
\end{equation}
with an energy-momentum tensor built only on the nonlinear electromagnetic fields (since we are only interested in black hole solutions to general relativity, which allows us to assume that the mass and charge of the system are only at its origin \footnote{In this work we are neglecting the baryonic contribution to the stress-energy tensor, though it generates the neutrinos we will make use of in order to explore some nonlinear theories of the electromagnetism, because we assume a situation in which it has already collapsed into a black hole (presumed to be charged). In this case, neutrinos could be treated as test-particles in this just formed black hole spacetime.}), given by \citep{2014PhLB..734..396P} 
\begin{equation}
4\pi T_{\mu\nu}\doteq\frac{2}{\sqrt[]{-g}}\frac{\partial L}{\partial g^{\mu\nu}}= 4L_FF_{\mu\alpha} F_{\nu\beta}g^{\alpha\beta}
-(L-GL_G)g_{\mu\nu}\, .
\label{d3}
\end{equation}
We have defined in the above equations that $L_X$ is the derivative of the Lagrangian density $L$ with respect to the invariant $X$, $F\doteq F^{\mu\nu}F_{\mu\nu}$, $G\doteq
\Fstar^{\mu\nu}F_{\mu\nu}$, $\Fstar^{\mu\nu} \doteq \eta^{\mu\nu\alpha\beta}
F_{\alpha\beta}/(2\sqrt{-g})$, $\eta^{0123}\doteq +1$, is a totally
antisymmetric tensor, $F_{\mu\nu}\doteq \partial_{\mu}A_{\nu}- \partial_{\nu}A_{\mu}$ the electromagnetic field four-tensor,
$A_{\mu}$ the electromagnetic four-potential and
$\Fstar^{\mu\nu}$ is its associate dual \cite{1975ctf..book.....L}. Besides,
$g$  has been defined as the determinant of the metric given
by Eq. (\ref{metric2}). In the above equations, only for mathematical convenience, we have taken $L=4\pi L_{Ga}$, where $L_{Ga}$ is the Lagrangian density in Gaussian units [for instance, for Maxwell's electromagnetism we have $L_{Ga}=-F/(16\pi)$].  Finally, let us define the electromagnetic fields by means of: $F_{tr}\doteq E_r$, $F_{t\theta}\doteq E_{\theta}$,
$F_{r\varphi}\doteq B_{\theta}$ and $F_{\varphi\theta}\doteq B_r$. Local fields are to be obtained by means of a tetrad decomposition of $F_{\mu\nu}$ following the above-mentioned definitions. 
Notice from the second term of Eq. (\ref{d2}) that we are
assuming our system is such that its current four-vector is null.

In the spherically symmetric case, the above field equations with asymptotically flat
black hole solutions lead to \cite{2014PhLB..734..396P}
\begin{equation}
g_{00}(r) = 1 - \frac{2M}{r} + \frac{2 Q A_0 }{r}-\frac{2 \cal{N}}{r}
\label{ee1}
\end{equation}
and
\begin{equation}
\frac{\partial L}{\partial E_{r0}}=  \frac{Q}{r^2}\label{ee2}
\end{equation}
with
\begin{equation}
E_{r0}\doteq - \frac{\partial A_0}{\partial r}\;\;\mbox{and}\;\; \frac{\partial
\cal{N}}{\partial r}\doteq - Lr^2 .
\label{ee3}
\end{equation}
[$L=L(F)$ in this case
since we are assuming the nonexistence of magnetic charges.]
The constants $M$ and $Q$ are the total mass (total energy) and charge of the system,
respectively, and are formally constants of integration. In Eq. (\ref{ee1}), a gauge has
been imposed such that $A_0(r)$ and ${\cal N}$ (the part of the total electromagnetic energy explicitly associated with a nonlinear Lagrangian) are null at infinity, guaranteeing the asymptotic flatness of the solutions and the asymptotic nullity of electromagnetic fields. Given $E_{r0}$ and $L=L(E_{r0})$, $A_{0}(r)$
and ${\cal N}(r)$ can be obtained by means of integration from an arbitrary radial
coordinate $r$ up to infinity.

Let us assume that the fields for the slowly rotating black holes are
\begin{eqnarray}
E_{r} & = & E_{r0}+{\cal O}\left({a^2} \right),\;\; B_r=B_{ra}a + {\cal O}
\left({a^2}\right), \; \;  \nonumber \\
E_{\theta} & = & {\cal O}\left({a^2} \right),\;\;B_{\theta}= B_{\theta a}a + {\cal
O} \left({a^2} \right)
\label{d4}.
\end{eqnarray}
By substituting Eqs. (\ref{metric2}) and (\ref{d4}) into Eq. (\ref{d2}), one can easily show that the only new equation arising, apart from the one in the spherically symmetric case, reads
\begin{eqnarray}
8 B_{\theta a} E_{r0} g_{00} L_F &+& 2 L A(r) \sin^2\theta=\sin^2\theta\left\{
\frac{A(r)(g_{00}r)'}{r^2}\right.\nonumber\\ &&\left. + \frac{1}{2} [g''_{00}A(r)-g_{00}A''(r)] \right\} ,
\label{d5}
\end{eqnarray}
where the prime symbol stands for the derivative with respect to the
$r$ coordinate. Since the Lagrangian $L(F,G)$ is an at least quadratic function of the
fields, then in the above equation it is implicit that $L$ and $L_F$ are evaluated at
$a=0$. From Eq. (\ref{d5}), one can immediately check that it is meaningful just
if
\begin{equation}
B_{\theta a}=f(r) \sin^2\theta ,
\label{d6a}
\end{equation}
where $f(r)$ is an arbitrary function of the radial coordinate. 
The equation governing the field components $B_{r a}$ and $B_{\theta a}$ can be obtained
from Eqs. (\ref{d2}) and (\ref{emid}) and are
\begin{equation}
\frac{\partial B_{ra}}{\partial r}+\frac{\partial B_{\theta a}}{\partial \theta}=0
\label{d7}
\end{equation}
and
\begin{eqnarray}
0&=&\sin\theta\frac{\partial}{\partial r} \left(L_F \left[-A(r)E_{r0} + g_{00}
\frac{B_{\theta a}} {\sin^2 \theta}\right]\right) \nonumber \\
&-& \frac{1}{r^2} \frac{\partial}{\partial\theta} \left(\frac{L_F B_{ra}}{\sin\theta}-
 \frac{r^2 L_G E_{r0}}{a}\right)  \; .
\label{d8}
\end{eqnarray}
From Eqs. (\ref{d6a}) and (\ref{d7}), it follows that
\begin{equation}
B_{r a}=g(r)\sin 2\theta\label{d8a},
\end{equation}
which leads to the very simple relation
\begin{equation}
f(r)=-g'(r)\label{d8b}.
\end{equation}
Since the Lagrangian must be an even power of the invariant $G$, it is straightforward to
see that
\begin{equation}
L_G=-\frac{8aL_YE_{r0}B_{ra}}{r^2\sin\theta} + {\cal O}\left(\frac{a^2}{r^2}\right),\;\;\;
 Y\doteq G^2\label{d9}.
\end{equation}
Finally, gathering Eqs. (\ref{d6a}), (\ref{d8a}) and  (\ref{d9}), Eq. (\ref{d8}) can be
cast in the form
\begin{equation}
-\{L_F[A(r)E_{r0}+g_{00}g'(r)]\}'+\frac{2g(r)}{r^2}[L_F+8L_YE^2_{r0}]=0\label{d10}.
\end{equation}
Up to zeroth order, one can also put Eq. (\ref{ee1}), with the help of Eq. (\ref{ee2}),
to the form
\begin{equation}
(g_{00}r)'=2 L r^2-2QE_{r0}+1\label{d10a}.
\end{equation}
Then, from Eqs. (\ref{d5}), (\ref{d6a}), (\ref{d8b}) and the above one, we simply have
\begin{equation}
2 Q g'(r) g_{00} = (-2 Q E_{r0} + 1)A(r) + \frac{r^2}{2} [g''_{00}A(r)-g_{00}A''(r)]
\label{d11}.
\end{equation}
Hence, we have two undetermined functions $g(r)$ and $A(r)$ and two coupled equations,
Eqs. (\ref{d10}) and (\ref{d11}). As the boundary condition, for large $r$, the functions should
approach their Maxwellian (Ma) counterparts,
\begin{eqnarray}
&&A(r)  \rightarrow  A(r)_{Ma}  =  g_{00}-1 = \frac{Q^2}{r^2}- \frac{2M}{r},\;\;\;
\nonumber \\
&&g(r)  \rightarrow  g(r)_{Ma} = \frac{Q}{r}
\label{d12},
\end{eqnarray}
as it can be shown by choosing $L(F,G)=-F/4$ and solving the equations for the metric
$g_{00}$, $E_{r0}$ and Eqs.~(\ref{d10}) and (\ref{d11}). 

Since in general the problem set out above can just be solved numerically, it turns out that the $r$ variable is not convenient for this end. Numerically, it is much more
suitable to use the dimensionless variable $u$, defined as $u\doteq M/r$. 
Concerning the integration of the aforesaid equations, $u$ should run from $0$ to $u_h$, where the latter is given as the smallest
solution to $g_{00}(u_h)=0$. The quantity $u_h$ is called the outermost horizon of the black hole and determining it is important because it defines the region of physical interest ($u<u_h$), which has an impact on all position-dependent observables.

\section{Geodesics in slowly rotating nonlinear spacetimes}
\label{geodesic}

Forasmuch as we are interested in describing neutrinos in spacetimes related to axially
symmetric
nonlinear black holes, the general study of geodesics is necessary, given that these
particles have no charge and hence are not sensitive to forces of electromagnetic origin. This is also so since
the Dirac equation in the limit of the WKB approximation (the one we shall be interested in here) assures that the phase part of neutrino spinors satisfy a Hamilton-Jacobi-like
equation \cite{2014arXiv1410.1523V} and by assuming that their amplitudes vary slowly, they do not play a role for convenient spacetime distances, revealing therefore the test-particle aspects of the neutrinos (quite similarly to the notion of rays in optics).
We underline from the previous sentences that we are overlooking the interaction of the B-field with the
neutrino anomalous magnetic moment and spin. Upon the aforesaid premises, there are several ways of describing neutrino trajectories. An elegant approach would be solving the associated
Hamilton-Jacobi equation \cite{1975ctf..book.....L} for the spacetime given by Eq.
(\ref{metric2}). Nevertheless, we will follow
the Lagrangian approach. Given that we are working up to first order in ``$a/r$'', due to the frame dragging effect, test-particles only remain confined in a plane if it is the equatorial one \cite{1995Kidder}, and thus for simplicity we limit our analysis to $\theta=\pi/2$. 
For this case, the proper Lagrangian for test
particles (t.p.) is \cite{1983mtbh.book.....C}
\begin{equation}
L^{(t.p.)}\doteq \frac{1}{2}\left[g_{00} \dot{t}^2-\frac{\dot{r}^2}{g_{00}}-r^2  \dot{\varphi}^2-2aA(r) \dot{t}\dot{\varphi}\right]
\label{lagtp},
\end{equation}
where $\dot{x}^{\mu}\doteq dx^{\mu}/d\lambda$, with $\lambda$ an affine parameter
along the
curve followed by the test particle. From Eq. (\ref{metric2}), the coordinates $t$ and
$\varphi$ are cyclic ones for the above Lagrangian. Hence, the quantities $p_t\doteq E$
and $p_{\varphi}\doteq -l$, with $p_{\mu}=g_{\mu\nu}p^{\mu}$, $p^{\mu}\doteq
m\dot{x}^{\mu}$, $m$ the rest mass of the test particle of interest, are constants along
the geodesics. 

From our previous definitions and Eq. (\ref{lagtp}), we thus have
\begin{equation}
\dot{t}=\frac{{\tilde E}r^2+aA(r) {\tilde l}} {g_{00}(r)r^2},\;\;\dot{\varphi} =\frac{{
\tilde l}g_{00}(r)-{\tilde E} aA(r)}{g_{00}(r) r^2} ,
\label{firstint}
\end{equation}
where we neglected terms of second order in $a/r$ and defined that for any quantity $C$, ${\tilde C}\doteq C/m$. Another first integral that comes {out of} our prescription is
\begin{equation}
\dot{r}^2={\tilde E}^2-g_{00}(r)\left[\frac{{\tilde l}^2} {r^2}-
\frac{2{\tilde E}{\tilde l}aA(r)} {g_{00}(r)r^2} +\epsilon\right] ,
\label{rdot}
\end{equation}
where $\epsilon =0,1$, according to which the geodesic is light-like or time-like,
 respectively. The above equation is obtained by means of the line element given
by Eq. (\ref{metric2}). 

Just for the sake of completeness, the last first integral of our analysis is $\dot{\theta}=0$. From Eq. (\ref{rdot}), one can even define an effective potential by means of ${\tilde V}^2 = {\tilde E}^2$, for $\dot{r}=0$, which then reads \cite{1973grav.book.....M,Pugliese2011-2013}
\begin{equation}
{\tilde V}_{\pm}=\frac{{\tilde l}aA(r)}{r^2}\pm \sqrt{g_{00}(r)\left[\frac{{
\tilde l}^2}{r^2} +\epsilon\right]} .
\label{effpot}
\end{equation}
One could just work with ${\tilde V}_{+}$, {since} the ``symmetry rule'' ${\tilde V}_{-}({\tilde l}) = -{\tilde V}_{+}(-{\tilde l})$ holds \cite{Pugliese2011-2013}. All features
characterizing the motion of neutral test particles can be obtained by means of the
scrutiny of the above equation. We shall not perform such an analysis here, for we are only
interested in neutrino spin-flip transitions and flavor oscillations.

In what follows, we shall discuss
neutrino oscillations ($\nu_a \longrightarrow \nu_b$) in connection to the
oscillation length and the transition probability. Analysis regarding spin-flip will be done in Sec. \ref{spinprec}.

\section{Neutrino flavor oscillation}
\label{flavoroscillation}

As stated previously, neutrino flavor oscillations would take place due to the fact that
neutrino flavor eigenstates $|\nu_\alpha\rangle$ are linear combinations of neutrino mass eigenstates $|\nu_j\rangle$
as (see e.g. \cite{1997PhRvD..56.1895F} and references therein)
\begin{equation}
|\nu_{\alpha}\rangle= U_{\alpha j}\exp[-i\Phi_j]|\nu_{j}\rangle ,
\label{neutreigenst}
\end{equation}
where repeated indexes are summed over. In the above equation, the $\alpha$ index stands for the neutrino flavor eigenstates, while the $j$ one stands for the masses eigenstates. The matrix $U_{\alpha j}$ is a unitary matrix that gives the mixing - level-crossing - between the flavor eigenstates and the mass eigenstates. Besides, $\Phi_j$ is the phase associated with the $j$\textit{th} mass eigenstate. For curved
spacetimes, $\Phi_j$ reads \cite{1997PhRvD..56.1895F}
\begin{equation}
\Phi_j= \int P_{(j)\mu}dx^{\mu}
\label{phaseosc},
\end{equation}
where $P_{(j)\mu}$ is just to indicate the four-momentum of the mass eigenstate $j$. We
shall assume that neutrinos just have two spin flavors. It is
well-known \cite{1986qmv1.book.....C} in this case that one can introduce a mixing
angle, $\Theta$, such that the transition probability from one flavor eigenstate $\alpha$
to
another $\beta$ reads
\begin{equation}
P(\nu_{\alpha}\rightarrow \nu_{\beta})= \sin^2(2\Theta)\sin^2\left(\frac{\Phi_{jk}}{2}
\right),
\label{transprob}
\end{equation}
where $\Phi_{jk}\doteq \Phi_j-\Phi_k$.
Whenever one is interested in neutrino propagation in spacetimes given by Eq.
(\ref{metric2}), after Eqs. (\ref{firstint}) and (\ref{rdot}) are
taken into account, for the case $\dot{r}\neq 0$, Eq. (\ref{phaseosc}) can be cast into the form
\begin{eqnarray}
&&\Phi_j = \int dr\frac{m_j\epsilon}{\dot{r}}=\nonumber\\&& m_j^2 \int \frac{\epsilon dr}{\sqrt{E^2 -
g_{00}(r) \left[\frac{l^2}{r^2}- \frac{2E laA(r)}{g_{00}(r)r^2} +
m_j^2\epsilon \right]}} .
\label{phaseoscj}
\end{eqnarray}

Note that Eq. (\ref{phaseoscj}) is exact and becomes zero for null geodesics. This is easily
understood by the fact that $p_{\mu}dx^{\mu}=g_{\mu\beta}p^{\beta}dx^{\mu}\propto ds^2$,
which is zero for null paths. 
Hence, when it is stated in the literature that null paths are taken into consideration, approximations are done such that in parts of the Eq. (\ref{phaseoscj}) properties of null geodesic are utilized. (In the standard treatment one assumes that $p_{\mu}=g_{\mu\nu}p^{\nu}$ is defined along a time-like geodesic, while the tangent four-vector to the trajectory $dx^{\mu}/d\lambda$ is taken to be null-like, resulting in a factor of $2$ when compared to the case both $p_{\mu}$ and $\dot{x}^{\mu}$ are defined along time-like geodesics. See \cite{Ren2010} for further details.)
In the nonlinear case, it is momentous to
bear in mind that photons do not follow null geodesics in their background spacetimes,
but in the so-called effective geometries (see e.g. \cite{2002IJMPA..17.4187N} and
references therein). Therefore, the distinction between massive particles and photons in our case is paramount.

From the second expression in Eq. (\ref{firstint}) we see that in general it is impossible to have $\dot{\varphi}=0$. Hence, \textit{pure} radial geodesics do not exist in axially symmetric spacetimes. The origin for so is the dragging of inertial frames, also known as Lense-Thirring effect
\cite{1973grav.book.....M}. 
Nevertheless, approximating $\dot{\varphi} \sim{\cal O}(a)$ is always possible if one assumes that $\tilde{l}\sim {\cal O}(a)$. For these particular (nearly radial) geodesics, the effects introduced by the nonlinearities of the Lagrangians are completely washed away, since $E\gg m$, and $0<g_{00}\leq 1$ outside the horizon. Hence, although the neutrino oscillation expression is that from special relativity in this case, the general relativistic effect of frame dragging still persists on their trajectories.

\subsection{Neutrino oscillation length}

Another important quantity that {arises} in the description of neutrino oscillations is
the oscillation length \cite{Ren2010,2012IJTP...51.1111R}. Basically it estimates the length over
which a
given neutrino has to travel for $\Phi_{jk}$ to change by $2\pi$.
Therefore, for talking about oscillation lengths, the proper relativistic covariant
distance is of importance \cite{1997PhRvD..55.7960C}. For the case of the spacetimes
described by Eq. (\ref{metric2}), from Eqs. (\ref{firstint}) and (\ref{rdot}) and assuming that the particles involved have the same energy $E$ and are such that $E\gg m_{j,k}$, it follows that
\begin{equation}
L_{osc}\doteq \frac{dl_{pr}}{d\Phi_{jk}/(2\pi)} \simeq \frac{2\pi
E}{\sqrt{g_{00}} (m^2_j-m^2_k)} ,
\label{osclength}
\end{equation}
where we assumed that $dl_{pr}$ is the infinitesimal proper distance, given by
\cite{1975ctf..book.....L}
\begin{equation}
dl^2_{pr}=\left(-g_{ij} + \frac{g_{0i}g_{0j}}{g_{00}} \right)dx^idx^j
\label{properlength},
\end{equation}
with $i,j=1,2,3$. If one wants to restore the conventional units, the right-hand side of
the equality in Eq.(\ref{osclength}) must be multiplied by $\hbar/c^3$.

Hence, from Eq. (\ref{osclength}) we learn that the oscillation length decreases whenever
$g_{00}$ increases. This is exactly the case for nonlinear charged black holes, when
compared to a Schwarzschild black hole. This means that when the black hole is charged,
neutrinos will tend to oscillate more than they would when it is not charged, for each
spacetime point (location). 

\section{Neutrino spin precession}
\label{spinprec}
In this section we summarize the main points about neutrino flavor spin precession,
also named neutrino flavor spin-flip, or neutrino-antineutrino oscillations
\cite{2006IJMPD..15.1017D}, and study them in the framework of the metric given by
Eq. (\ref{metric2}). For point-like particles, the equations governing the spin $S^{\mu}$
coupling of test particles with the gravitational field are
\cite{1972gcpa.book.....W}
\begin{equation}
\frac{DS^{\mu}}{d\lambda}=0,\;\; \frac{Du^{\mu}}{d\lambda}=0 ,
\label{spineq}
\end{equation}
where $D/d\lambda$ stands for the absolute derivative with $\lambda$ an affine parameter \cite{1975ctf..book.....L}.
From the definition of the absolute derivative, one sees that the spin does
change whenever spin connections are not null, as contrary to the case of an
intrinsically flat Minkowski spacetime.

A proper analysis about the spin evolution
of a test particle by a local observer is done with the use of  (orthonormal) tetrads ($e^{a}_{\mu}$), i.e., \cite{1975ctf..book.....L}
\begin{equation}
g_{\mu\nu}= \eta_{ab}e^{a}_{\mu}e^{b}_{\nu},\;\;\; e^{a}_{\mu}e_{a}^{\nu}=
\delta^{\nu}{}_{\mu},\;\;\;e^{a}_{\mu}e_{b}^{\mu}=\delta^{a}_{b} ,
\label{tetradprop1}
\end{equation}
where
\begin{equation}
\eta_{ab}=diag(1,-1,-1,-1),\;\;\; e^{\mu a}=g^{\mu\nu}e^{a}_{\nu},\;\;\;
e^{a}_{\mu}=\eta^{a b} e_{b\mu} .
\label{tetradprop2}
\end{equation}
The tetrad decomposition of a four-vector $C^{\mu}$ is defined as $C^{a}\doteq e^{a}_{\mu}
C^{\mu}$, with the derivative of $C_b$, $C_{b,c}$, instead is defined by
\cite{1975ctf..book.....L}
\begin{equation}
C_{b,c}\doteq e^{\mu}_c \frac{\partial C_b}{\partial x^{\mu}} .
\label{tetder}
\end{equation}
From $u^a\doteq e^{a}_{\mu}u^{\mu}$, with $u^{\mu}\doteq dx^{\mu}/d\tau$, (the four-velocity of the test particle), it follows that $S^a_{\;\; ,b} u^b=dS^{a}/d\tau$, which allows
to conclude that $S^a_{\;\; ,b} =\partial S^a/\partial y^b$ and $u^a=dy^a/d\tau$, with $y^a$ the
coordinates utilized by the local observers. Hence, $e^{a}_{\mu}=\partial y^{a}/\partial
x^{\mu}$ and $ds^2=\eta_{ab}dy^a dy^b=g_{\mu\nu} dx^{\mu}dx^{\nu}$.

From Ref. \cite{1975ctf..book.....L}
\begin{equation}
A_{\mu;\nu}e^{\mu}_a e^{\nu}_b= A_{a,b}+\eta^{cd}\gamma_{cab} A_d ,
\label{tetcovder}
\end{equation}
where $\gamma_{abc}$ are the Ricci rotation coefficients, defined as
\cite{1975ctf..book.....L}
\begin{equation}
\gamma_{abc}\doteq e_{a\mu;\nu}e^{\mu}_be^{\nu}_c ,
\label{gammadef}
\end{equation}
and by using Eqs. (\ref{tetcovder}) and (\ref{gammadef}), Eq. (\ref{spineq}) can be
recast as (see for instance \cite{2006IJMPD..15.1017D})
\begin{equation}
\frac{dS^a}{d\tau} = \varpi^{ab} S_b,\;\; \frac{du^a}{d\tau}=\varpi^{ab}u_b ,
\label{spinteteq}
\end{equation}
with
\begin{equation}
\varpi^{ab}\doteq \eta^{ac}\eta^{bd}\gamma_{cde}u^e\label{gdef}.
\end{equation}
In virtue of the antisymmetry of $\gamma_{abc}$ on its first two indexes
\cite{1975ctf..book.....L}, it follows that $\varpi^{ab}$ is an antisymmetric tensor.
Hence, it can be decomposed into ``electric'' and ``magnetic'' parts, $E_i^{\varpi}$ and $B_i^{\varpi}$ respectively, quite similarly to what is done for the electromagnetic tensor. In other words,
\begin{equation}
E_i^{\varpi}=\varpi_{0i},\;\; \varpi_{ij}=-\epsilon_{ijk}B_k^{\varpi}\label{elemagnfields},
\end{equation}
with $\epsilon_{ijk}$ being a totally antisymmetric tensor such that $\epsilon_{123}
\doteq 1$. We have used the convention that the Latin indexes that run from zero to three
are the ones from the beginning of the alphabet ($a, b, c, ...$), while those that run
from one to three are the ones from the middle of the alphabet ($i, j, k, ...$).

In general, a neutrino has a nonzero velocity with respect to a tetrad defined at a given
point of the spacetime. Thence, it can always be defined a ``locally
comoving frame'', where in the latter it is instantaneously unmoving. In this
frame, ${\bar s}^ a\doteq \xi^ i \delta^ a_ i$ and $\bar{u}^ a=\delta^a_0$. Thus, by using the
Lorentz transformations to connect both systems, one ends up with the relation
\cite{2006IJMPD..15.1017D}
\begin{equation}
S^ a=\left[\vec{\xi}\cdot \vec{u},\vec{\xi}+
\frac{\vec{u}(\vec{\xi}\cdot \vec{u})}{1+u^ 0} \right] ,
\label{spininsyst}
\end{equation}
where $u^0$ and $\vec{u}$ are the temporal and the spatial components, respectively, of
the comoving frame with respect to the inertial one, or the four-velocity of the particle
in this reference system. By substituting Eq. (\ref{spininsyst}) in Eq. (\ref{spinteteq})
[it is important to use both equations], and taking it into account Eq.
(\ref{elemagnfields}), one arrives at \cite{2006IJMPD..15.1017D}
\begin{equation}
\frac{d\vec{\xi}}{d\tau}= \vec{\xi}\times \vec{\varpi},\;\;\vec{\varpi}\doteq \left(\vec{B}^{\varpi}
+ \frac{\vec{E}^{\varpi}\times \vec{u}}{1+u^0}\right)
\label{spinevol}.
\end{equation}
Then, it is an elementary task to verify that the spin $\vec{\xi}$ of the particle precesses about the vector $\vec{\varpi}$.

From now on, we shall be interested in applying the above formalism for the case of the
intrinsic (quantum) spin of neutrinos moving in spacetimes given by Eq. (\ref{metric2}).
To start with, as suggested by Eq. (\ref{metric2}), for local measurements, we
choose the tetrad
\begin{eqnarray}
e^{0}_{\mu} & = &
\left(\sqrt{g_{00}},0,0,-\frac{aA(r)\sin^2\theta}{\sqrt{g_{00}}}\right),\; e^{1}_{\mu} =
\left(0,\frac{1} {\sqrt{g_{00}}},0,0\right), \nonumber \\
\; e^{2}_{\mu} & = & (0,0,r,0),\; e^{3}_{\mu}=(0,0,0,r\sin\theta)
\label{tetrad}.
\end{eqnarray}
Just for the sake of completeness, the corresponding inverse tetrad reads
\begin{eqnarray}
e^{\mu}_0 & = & \left(\frac{1}{\sqrt{g_{00}}},0,0,0\right), \;e^{\mu}_1 =
\left(0,\sqrt{g_{00}},0,0\right) , \\
e^{\mu}_2 & = & \left(0,0,\frac{1}{r},0 \right), \;e^{\mu}_3=\left(\frac{aA(r)\sin
\theta}{rg_{00}(r)},0,0,\frac{1}{r\sin\theta} \right)\nonumber
\label{tetrinv}.
\end{eqnarray}
It can be easily checked that the properties given by Eq.~(\ref{tetradprop1}) hold for
the above tetrad up to the first order in ``$a/r$'', as internal consistency demands. For the aforesaid tetrad, we now present the nonvanishing Ricci rotation coefficients for Eq.
(\ref{metric2}). They follow from Eq. (\ref{gammadef}) as
\begin{align}
\gamma_{010}&=-\frac{g_{00,r}}{2\sqrt{g_{00}}},& \gamma_{013}&=\frac{a\sin\theta[g_{00}A(r)_{,\,r} - A(r)
g_{00,\,r}]}{2r g_{00}},\nonumber \\
\gamma_{023}&=\frac{aA(r)\cos\theta}{r^2\sqrt{g_{00}}},& \gamma_{031}&=-\gamma_{013},
\nonumber\\
\gamma_{032}&=-\gamma_{023},& \gamma_{122}&=-\frac{\sqrt{g_{00}}}{r},
\label{riccirotcoeff} \\
\gamma_{130}&=-\gamma_{013},& \gamma_{133}&=\gamma_{122},\nonumber\\
\gamma_{230}&=-\gamma_{023},& \gamma_{233}&=-\frac{\cos\theta}{r\sin\theta} \;. \nonumber
\end{align}
In the above equations we have defined $ C_{,\,r} \doteq \partial C/\partial r$ for
a given function $C(r)$. From the geodesic motion of test particles, it follows that
\begin{equation}
u^a=\left[ \sqrt{g_{00}}\dot{t} -\frac{aA(r) \sin^2\theta}{\sqrt{g_{00}}} \dot{\varphi},
\frac{\dot{r}}{\sqrt{g_{00}}}, \dot{\theta}r,r\sin\theta\dot{\varphi} \right] .
\label{tetvel}
\end{equation}
From Eqs. (\ref{spinevol}), (\ref{elemagnfields}), (\ref{gdef}) and (\ref{riccirotcoeff}),
and taking {that the orbits lie in the plane}  $\theta =\pi/2 $, so that $\dot{\theta}=0$,
one obtains the ``electric'' component of tensor $\varpi_{ab}$ in the form
\begin{eqnarray}
\vec{E}^{\varpi} &=& \left[-\frac{g_{00,\,r} \dot{t}}{2} + \frac{aA(r)_{,\,r} 
\dot{\varphi}}{2},0, \right.\nonumber\\ &&\left.
\frac{a[A(r) g_{00,\,r} -g_{00}A(r)_{,\,r}]\dot{r} }{2rg^{\frac{3}{2}}_{00} }
\right] ,
\label{electricfield}
\end{eqnarray}
and the ``magnetic'' component
\begin{equation}
\vec{B}^{\varpi}=\left[0,-\sqrt{g_{00}} \dot{\varphi} + \frac{a[A(r)g_{00,\,r} -g_{00}A(r)_{,\,r}]\dot{t}}{2r \sqrt{g_{00}}}, 0 \right].
\label{magneticfield}
\end{equation}
Let us investigate now circular orbits under the condition that
$\dot{r}=0$ and $\partial{\tilde V}/\partial r=0$. From the critical points of the
effective potential ${\tilde V}$ for $\epsilon=1$, one has that it implies
\begin{equation}
{\tilde l}^2_{\pm}=\frac{g_{00,\,r}r^3}{2g_{00}-g_{00,\,r}r} \pm \frac{{\cal B} a}{(2g_{00}-g_{00,\,r}r)^2},
\label{angmomcirc}
\end{equation}
with 
\begin{equation}
{\cal B}=2\sqrt{2}\sqrt{r^7g_{00}^2g_{00,\,r}(2A-rA_{,r})^2}\label{circular_orbit_l_a}
\end{equation}
The remainder first integrals (\ref{firstint}) can be obtained from the above equations and the consideration that $\tilde{E}_{\pm}=\tilde{V}_{\pm}$.

For circular orbits in the equatorial plane, Eq. (\ref{spinevol}) tells us that the angular velocity of the precession is generically given by $|\vec{\varpi}|\hat\theta$. Therefore, we have that the spin-flip (s.f.) probability for neutrinos in a slowly rotating and charged spacetime is
\begin{equation}
{\cal P}_{s.f.}(\tau)=\sin^2\left({|\vec{\varpi}|\tau}\right)\label{probsf} \, .
\end{equation}
We recall that in the above equation it is assumed that the spin of the neutrino is initially anti-parallel to its momentum vector, as it is the case for left-handed ({\sl Dirac}) neutrinos.

\section{Neutrino oscillations and spin-flip for the Born-Infeld Lagrangian}
\label{nosfbi}

We now study neutrino spin-flip and neutrino oscillations for the Lagrangian density put forward by
Born and Infeld in the 1930's. It can be written as \cite{1934RSPSA.144..425B}
\begin{equation}
L_{B.I}=b^2\left[1-\sqrt{1+\frac{F}{2b^2}-\frac{G^2}{16b^4}} \right]\label{bilagrang}.
\end{equation}
In the above Lagrangian,  $b$ represents the scale field and it sets out the
upper limit
for the electric field when magnetic aspects do not take place.
It was recently shown \cite{2006PhRvL..96c0402C,2011PhLA..375.1391F} that the $b$ proposed by Born and Infeld
is not able to reproduce the energy spectrum of the hydrogen atom, both in the frameworks of
nonrelativistic and relativistic quantum mechanics. A value
much larger than that one predicted under the unitary viewpoint is required, although a
definite one has not been obtained. This fact makes the direct probe of the Born-Infeld
Lagrangian even subtler, due to the present difficulty in getting hyper-high
electromagnetic fields in laboratory.
Apart from
the aforementioned problematic issue, hereafter we treat such a scale field as a
free parameter.

We start out our analyzes with the behavior of the metric given by Eq. (\ref{metric2}) and the electromagnetic fields for a slowly rotating axially symmetric spacetime in the scope of
the Born-Infeld Lagrangian. Such an analysis is important for it would give the range of the
parameters where considerable departures from the static nonlinear counterpart could take
place. To this end, we note that it is already known that the Born-Infeld Lagrangian leads to an exact solution to Einstein's equations in the spherically symmetric case
\cite{2002CQGra..19..601B} (the seed for slowly rotating analyses), and it can be cast as
\begin{eqnarray}
&&g_{00} = 1- 2u + \frac{2}{3u^2}{(bM)}^2\left(1-
\sqrt{1+ \frac{\alpha^2u^4}{{(bM)}^2}} \right)\nonumber\\ &+& \frac{2\alpha^2u}{3}
\sqrt{\frac{{bM}}{|\alpha|}} {\cal F}\left[ \arccos\left({\frac{{bM}-
|\alpha|u^2}{{bM} + |\alpha|u^2}}\right),\frac{1}{\sqrt{2}}\right]
\label{big00},
\end{eqnarray}
where ${\cal F}[...,1/\sqrt{2}]$ is the elliptic function of first kind \cite{2007tisp.book.....G}.

\begin{figure}[!hbt]
\centering
\includegraphics[width=\hsize,clip]{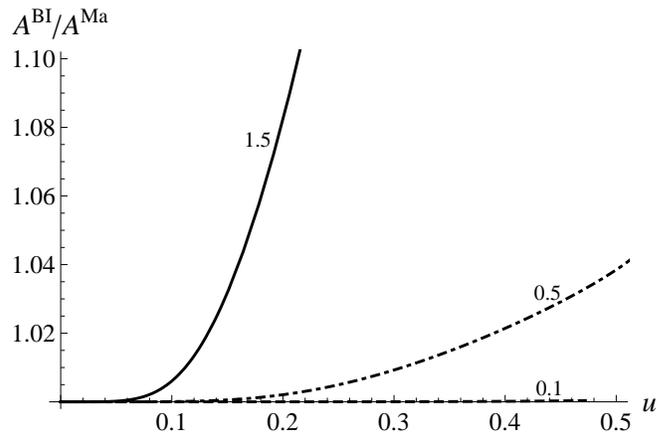}
\caption{{\small\sf Ratio of the off-diagonal term $A(r)$ in Eq. (\ref{metric2}) coming from Born-Infeld (BI) Lagrangian and the Maxwell (Ma) Lagrangian for selected values of the parameter $\alpha$ with $bM=0.017$. The value of $bM$ was chosen such that it is in agreement with the condition $bM>10^{-4}$, valid for astrophysical bodies with some solar masses, and Eq. (\ref{biineq}) is satisfied for all selected $\alpha$. In this case, the associated black holes do have just one horizon and do not have classical counterparts.}}
\label{metricratio}
\end{figure}

\begin{figure}[!hbt]
\centering
\includegraphics[width=\hsize,clip]{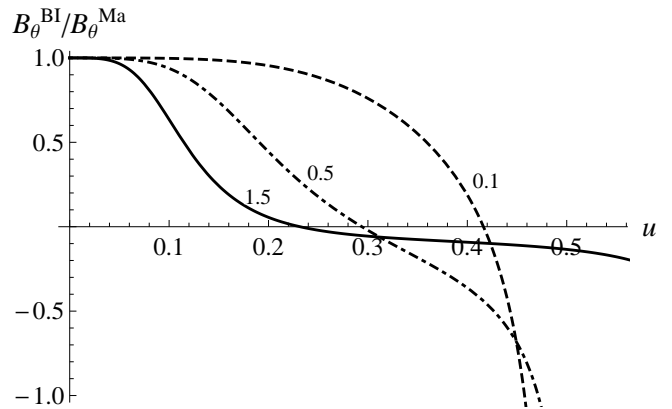}
\caption{{\small\sf Ratio of the polar magnetic fields for the same theories,
selected values of $\alpha$ and meaning of the curves as
in Fig. \ref{metricratio}.}}
\label{polarratio}
\end{figure}

\begin{figure}[!hbt]
\centering
\includegraphics[width=\hsize,clip]{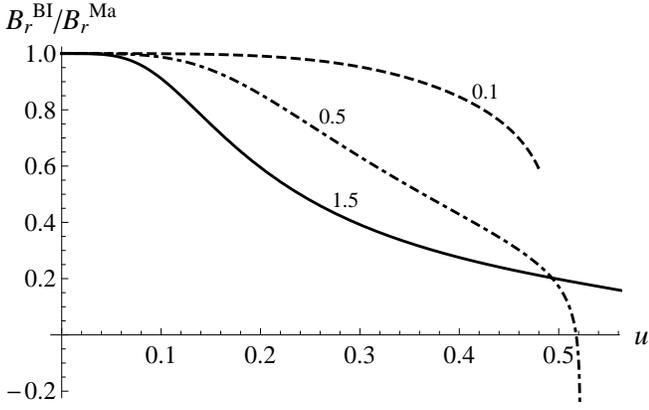}
\caption{{\small\sf Ratio of radial components of the magnetic field for the same
assumptions of Figs. \ref{metricratio} and \ref{polarratio}. }}
\label{radialratio}
\end{figure}

In Figs. \ref{metricratio}, \ref{polarratio} and \ref{radialratio} we show
the numerical integration of Eqs. (\ref{d10}) and (\ref{d11}) in terms of the dimensionless variable $u$ for some selected values of $\alpha\doteq Q/M$, with $bM=0.017$, for the components of the polar and radial magnetic fields and the metric functions $A(r)$, vis-\`a-vis the Maxwellian Lagrangian. The motivations for values of $\alpha$ of order of unity are given in Sec. X. We now justify the value of $bM$ picked out. We are working with geometric units, which means that $b$ has units of the inverse of length while $M$ has units of length, rendering thus $bM$ dimensionless. Their conversion to cgs units is done by means of the following rules \cite{2014PhLB..734..396P,Wald-book}: $M$[g]=$M$[cm]c$^2/G$ and $b$[statvolt/cm]=b[cm$^{-1}]c^2/\sqrt[]{G}$.
Let us assume that we work with black holes of around 3 solar masses, as it seems reasonable for those having a relationship with neutron stars. This means that in geometric units, $M\approx 5 \times 10^5$ cm. We know experimentally that $b$ must be larger than $10^{15}$ statvolt/cm in order to explain the observed energy levels of the hydrogen atom \cite{2006PhRvL..96c0402C,2011PhLA..375.1391F}. Therefore, in geometric units, $b>10^{-10}$cm$^{-1}$. This finally means that $bM> 10^{-4}$ is very reasonable for astrophysical black holes, the ones we will be interested in here.
One can perceive from the plots that the distinctness between theories starts to become more accentuated the closer the horizon is approached for each $\alpha$. Near that border there seems to exist a region where the magnetic field experiences a sharp deviation w.r.t. Maxwellian one.
Note that for all $\alpha$ selected the value of $bM$ also satisfies
\begin{equation}
bM< \frac{9}{|\alpha|^3\,{\cal F}^2\left[\pi,\,\frac{1}{\sqrt{2}}\right]}\approx \frac{0.654}{|
\alpha|^3}\label{biineq}.
\end{equation}
This means that the associated black holes just exhibit one non-degenerated  horizon \cite{1994CQGra..11.1469D,2011PhRvD..83j4009E}. {Consequently, $g_{00}$ is a monotonic function of the radial
coordinate.} We recall that Eq. (\ref{biineq}) does not have a classical limit, formally obtained when $b$ tends to infinity. {Whenever the inequality in Eq. (\ref{biineq}) occurs,} one should expect significant deviations from the standard classical solution, as it can be seen again in Figs. \ref{metricratio}, \ref{polarratio} and \ref{radialratio} {for some values of the parameter $\alpha$}. For the case where Eq. (\ref{biineq}) is not valid, Einstein-Born-Infeld black holes are the generalization of their Einstein-Maxwell counterparts. Naturally, when naked singularities are present, the aforementioned solutions may be considerably different, especially close to the singularity. There the fields coming from Born-Infeld Lagrangian are minute when compared to their associated classical ones, due to the regularity of the former Lagrangian at the singularity.

We emphasize that in the light of the black hole energy decomposition in nonlinear electrodynamics \cite{2014PhLB..734..396P}, when it applies, the comparison of a nonlinear black hole with its linear counterpart (Maxwellian Lagrangian) at the same value of $M, Q$ and $a$ generally means black holes with different irreducible masses \cite{Christodoulou1970-1971}.

\begin{figure}[!hbt]
\centering
\includegraphics[width=\hsize,clip]{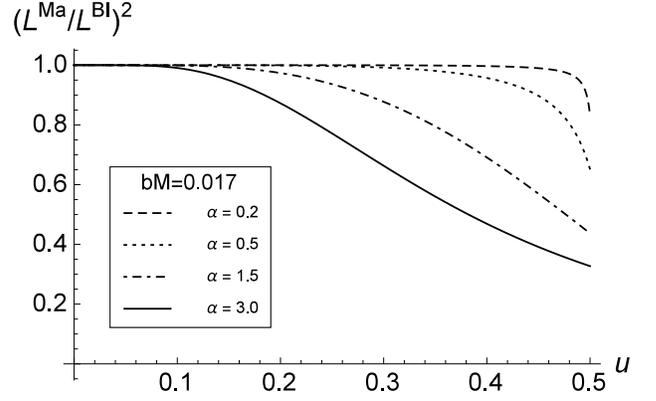}
\caption{{\small\sf Maxwell to Born-Infeld black holes oscillation lengths ratio for selected values of $\alpha$ for a fixed value of $bM$ satisfying $bM>10^{-4}$ and Eq. (\ref{biineq}). }}
\label{lengthratio}
\end{figure}

We now progress on with neutrino oscillations analyses within the Born-Infeld Lagrangian. We primarily want to compute the oscillation length when just two neutrino flavors are considered. This is easily accomplished with the use of Eqs. (\ref{osclength}) and (\ref{big00}). In Fig. \ref{lengthratio}, we plot the ratio of the oscillation lengths for selected values of $\alpha$, the charge-to-mass ratio, with a fixed value of $bM$ that satisfies $bM>10^{-4}$ and Eq. (\ref{biineq}), assuming that the oscillating particles do have the same energy $E$. Notice that in all cases the neutrino oscillation lengths are smaller in the scope of the Maxwellian Lagrangian. This can be physically understood due to the nonlinearities brought in by the Born-Infeld Lagrangian. Either theory, though, leads to smaller neutrino oscillation lengths than the ones for Schwarzschild black holes.

Let us now take a closer look at the spin-flip for the Born-Infeld Lagrangian. We limit ourselves to circular orbits. 
Given that in this case the frequency of spin-flip for slowly rotating spacetimes should generically read $\vec{\varpi}=(\varpi^{s.f.}_{sph.} +\Delta\varpi^{s.f.})\hat{\theta}$, with $|\Delta \varpi^{s.f.}_{sph.}|/|\varpi^{s.f.}|\ll 1$, we start out our analyses with the dominant spherically symmetric case. From Eqs. (\ref{tetvel}), (\ref{electricfield}), (\ref{magneticfield}), (\ref{angmomcirc}) and the associate first integrals (\ref{firstint}), we have that the angular velocity of precession of the neutrino spin, Eq. (\ref{spinevol}), can be simplified up to a sign to
\begin{equation}
\vec{\varpi}^{s.f.}_{sph.}=  \hat\theta \sqrt{\frac{g_{00,\,r}}{2r}} .
\label{Gcircspher}
\end{equation}
{We highlight that the main facets of the frequency of spin-flip depend upon the choice of the parameter $bM$. Whenever $bM\gg1$,
Eq. (\ref{big00}) gives us
\begin{equation}
g_{00}= 1-2u+\alpha^2u^2-\frac{\alpha ^4 u^6}{20 (bM)^2} + {\cal O}\left[\frac{1}{(bM)^3}\right]\label{big00bMgg1}.
\end{equation}
This signifies that the Einstein-Born-Infeld theory leads to the lessening of the metric when compared to the Reissner-Nordstr\"{o}m metric.
Therefore, $bM\gg1$ leads to an augment of the frequency of spin-flip, Eq. (\ref{Gcircspher}), when compared to the classical case.
One also perceives from Eq. (\ref{big00bMgg1}) that, like in the classical case, $\varpi^{s.f.}$ diminishes with the increase of $\alpha$.

Whenever $bM\ll1$, we have that Eq. (\ref{big00}) can be approximated to
\begin{equation}
g_{00}=1-2u + \frac{4}{3} \alpha ^{3/2} \sqrt{bM} {\cal F}\left[\pi,\frac{1}{\sqrt{2}}\right] u +{\cal O}(bM) \label{big00bMll1}.
\end{equation}
The comparison of the case $bM\ll 1$ with the Reissner-Nordstr\"om solution (same $\alpha$) is not immediate, though. For a given $\alpha$, if $u\leq {\cal F}[\pi, 1/\sqrt{2}] \sqrt{bM/\alpha}/3$, then it can be shown that $\varpi^{s.f.}_{bM\ll1}\geq \varpi^{s.f.}_{Ma}$. For a given $u$, the frequency of spin-flip increases with the decrease of $\alpha$. Notwithstanding, either if $bM\ll 1$ or $bM\gg 1$, the frequency of spin-flip for the case the charge is absent is larger than the case it is not. We exemplify the aforementioned scenario in Fig. \ref{sinspinfrequencybMll1}.
\begin{figure}[!hbt]
\centering
\includegraphics[width=\hsize,clip]{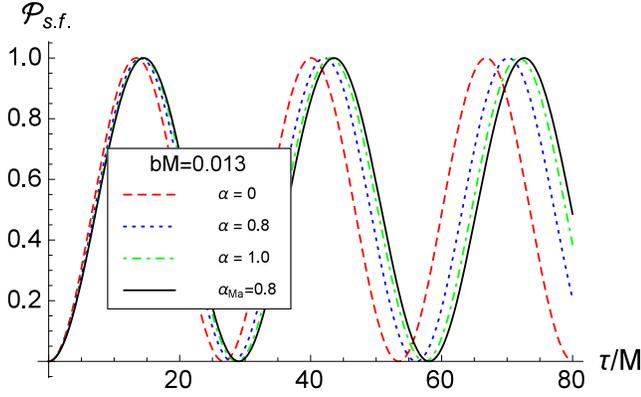}
\caption{{\small\sf Spherically symmetric transition probability of neutrino spin-flip, Eqs.
(\ref{probsf}) and (\ref{Gcircspher}), for selected values of $\alpha$ and $bM$, for circular orbits at
$u=0.24$. Notice that in this case, $u\leq {\cal F}[\pi, 1/\sqrt{2}] \sqrt{bM/\alpha}/3$, and so the spin-flip frequency in the Born-Infeld theory is larger than its Maxwellian counterpart. We point out that in this case the electromagnetic theories for a given $\alpha$ would differ after $\tau \approx 20 M$, which for stars with some solar masses would be equivalent to $(10^{-4}-10^{-3})$ s.}}
\label{sinspinfrequencybMll1}
\end{figure}

Now we investigate the changes impinged on $\vec{\varpi}$ due to the Born-Infeld nonlinearities and the slow rotation of the spacetime ($\Delta \varpi^{s.f.}_{BI}$). This is more readily  understood when compared to its Maxwellian counterpart. Figure \ref{deltaspinflip} shows the numerical analysis for circular orbits in the Born-Infeld Lagrangian for $bM=0.013$ and some choices of the parameter $\alpha$ (for specificity we have chosen here $\tilde{l}_+>0$ and $\tilde{E}_+$). Notice that the Maxwellian corrections to $\vec{\varpi}$ are always smaller (in modulus) than their Born-Infeld counterparts for a given $a$. This means that the Born-Infeld theory induces faster neutrino-antineutrino changes than the Maxwellian one when only small rotations are concerned.

\begin{figure}[!hbt]
\centering
\includegraphics[width=\hsize,clip]{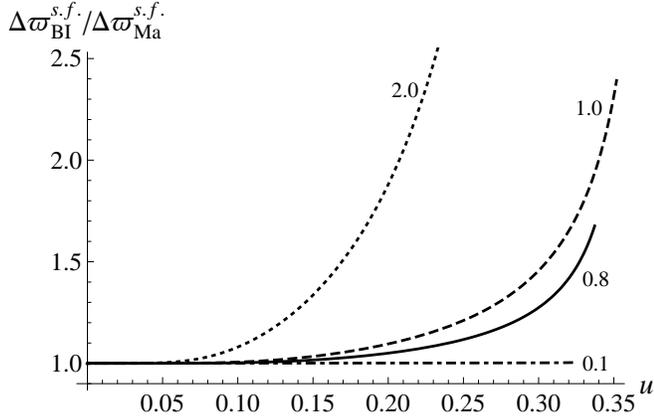}
\caption{{\small\sf Induced spin-flip frequencies due to slow rotation for the Born-Infeld theory when compared to its Maxwellian counterpart for $bM=0.013$ and selected values of $\alpha$ within the context of circular orbits. The Born-Infeld frequency induction due to slow rotation is always larger in magnitude than the classical case.}}
\label{deltaspinflip}
\end{figure}

\section{Nonlinear precessions}
\label{nonlinprec}
Next, we shall deduce the angular velocity of precession of gyroscopes placed at a point of the spacetime defined by Eq. (\ref{metric2}). The formalism is the same as the one for spin-flip described previously. Now, however, we place particles at rest (with respect to local observers) at given spacetime points and their precession is uniquely due to the ``rotation of the spacetime''. This is nothing more than the effect of dragging of inertial frames \cite{1973grav.book.....M}. It could be seen as another effect to probe Eq. (\ref{metric2}) in the context of nonlinear theories of the electromagnetism.

As it can be seen in Refs.  \cite{1973grav.book.....M,2006IJMPD..15.1017D} and
from Eq. (\ref{spineq}) when it is expanded in terms of connections, the components of the angular velocity $\Omega^{k}$ of precession of a gyroscope with respect to a given tetrad can generically be calculated by means of the relation \cite{1973grav.book.....M}
\begin{equation}
\epsilon_{ijk}\Omega^{k}=-\Gamma_{ij0}
\label{d16}
\end{equation}
where the tetrad decomposition of the Christoffel symbol is defined by the expression
\begin{equation}
\Gamma_{ijk} = e^{\mu}_{i}e^{\nu}_{j}e^{\beta}_{k}\Gamma_{\mu\nu\beta},\;\;
\Gamma_{\mu\nu\beta}\doteq \frac{1}{2}(\partial_{\beta}g_{\mu\nu} +
\partial_{\nu}g_{\mu\beta} -\partial_{\mu}g_{\nu\beta}) .
\label{d17}
\end{equation}
Notice that the sign present in Eq. (\ref{d16}) does not appear in Ref.
\cite{1973grav.book.....M} due to fact that we chose a different signature to the
metric.

Subsequently to uninvolved calculations one obtains the following results for the $\Omega^{k}$
components of the metric related to Eq. (\ref{metric2}) and the tetrad given by Eq.
(\ref{tetrad})
\begin{eqnarray}
\Omega^{r} &=& -\frac{a A(r)\cos\theta}{\sqrt{g_{00}}r^2},\nonumber  \\
\Omega^{\theta} &=& \frac{a [g_{00}\partial_{r}A(r)-A(r)\partial_{r}g_{00}]
\sin\theta}{2g_{00}r} \label{d18} \\
\Omega^{\phi} &=& 0. \nonumber
\end{eqnarray}
As we have already advanced, the above local angular precession can also be got (apart
from a sign {due to the vector product order chosen in Ref. \cite{1973grav.book.....M}}) from the spin-flip formalism by assuming there $u^a=\delta^{a}_0$. 

In Fig. \ref{girr} we plot the Born-Infeld to Maxwell ratio of the radial precession frequency, as appearing in Eq.~(\ref{d18}), for various values of $\alpha$ with $bM=0.1$ and an arbitrary $\theta$. One sees that there is a minute change of $\Omega^{r}$ coming from the aforementioned theories. In Fig. \ref{girth}, the plot illustrates the polar angular
frequency component in Eq.~(\ref{d18}). For {this case} the Born-Infeld theory could deviate considerably from the Maxwell one. This is particularly the case for large values of $\alpha$ and distances close to their associated outermost horizons.
\begin{figure}[!hbt]
\centering
\includegraphics[width=\hsize,clip]{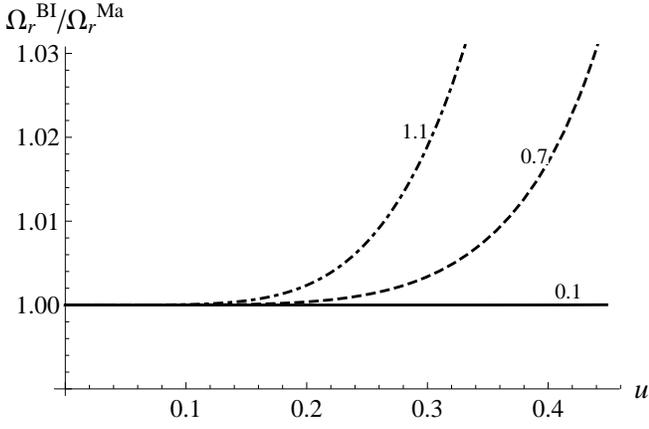}
\caption{{\small\sf Born-Infeld to Maxwell ratio of $\Omega^r$ appearing
in Eq.~(\ref{d18}) for various values of $\alpha$ and $\theta$, with
$bM=0.1$.}}
\label{girr}
\end{figure}
\begin{figure}[!hbt]
\centering
\includegraphics[width=\hsize,clip]{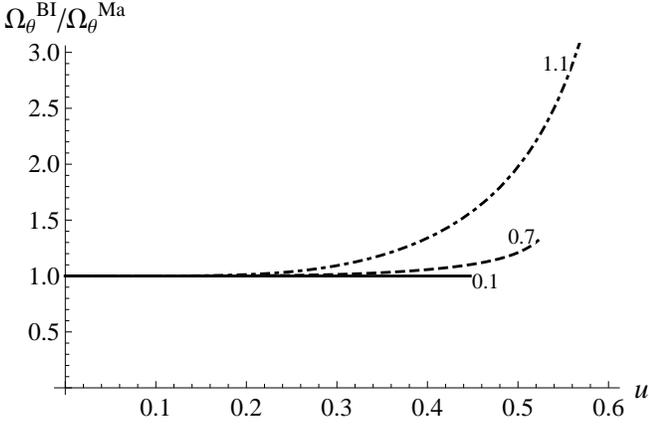}
\caption{{\small\sf Born-Infeld to Maxwell ratio of the polar frequency
component $\Omega^{\theta}$ appearing in Eq. (\ref{d18}) for various values of $\alpha$
and $\theta$, with $bM=0.1$.}}
\label{girth}
\end{figure}

Hence, if measurements could be done concerning the polar component of the precession of gyroscope-like systems (such as planets) in the environs of the horizon of a slowly rotating black hole (where we expect the precession should be more relevant), then one would be directly probing intrinsic properties of such spacetime, as well as of electromagnetism, this way overcoming the current experimental difficulty of probing it on terrestrial and atmospheric laboratories.

\section{r-process in supernova events around RNCBH spacetimes}
\label{rproc}
In what follows we shall revisit the effects of gravity on the
energy spectra of neutrinos (${\nu}_e$) and antineutrinos ($\overline{\nu}_e$) outflowing from the very inner ejecta of a type II supernova explosion {when a RNCBH
might already have been formed there}.

In so doing, we follow inasmuch the seminal paper by Fuller and Qian
\cite{1996NuPhA.606..167F} on the astrophysics of neutrinos escaping from the gravitational field a proto-neutron star, since except for the specific strength of the gravitational field both spacetimes are rather similar, as we already pointed out above.

In such environments neutrinos are subjected to the strongest gravitational redshift induced by any astrophysical object. 
Now, since the energy of electron antineutrinos is higher than their electron partners ($\langle E_{\overline{\nu}_e}\rangle > \langle E_{{\nu}_e}\rangle$), the former decouple deeper in the gravitational potential well of a putative RNCBH than the latter do.
Then, the $\overline{\nu}_e$ undergo the larger gravity-induced redshift action as compared to ${\nu}_e$, and this effect should manifest itself in several ways in the dynamics of the supernova explosion. It is then expected that such a differential redshift modifies the electron fraction ($Y_e$), which is defined as
\begin{equation}
Y_e \simeq \left[ 1 + \frac{ S_{\overline{\nu}_e p} }{S_{{\nu}_e n} } \right]^{-1} \simeq
\left[1 + \frac{ L_{\overline{\nu}_e} \, \langle E_{\overline{\nu}_e} \rangle}{ L_{\nu_e}
\, \langle E_{\nu_e}\rangle}  \right]^{-1} \; ,
\label{e-fraction}
\end{equation}
 that directly measures the neutron-to-proton ratio ($\frac{n}{p} \equiv \frac{1}{Y_e} - 1$) in the neutrino-driven supernova ejecta. This quantity is essential for any r-process nucleosynthesis occurring in this environment, which otherwise demand a low $Y_e$.

In this respect, the neutrino
${\nu}_e$ $\longrightarrow$ antineutrino $\overline{\nu}_e$ oscillations mediated by the 
gravitational collapse of the supernova inner core could properly explain the abnormally large abundance of
neutrons so as to support the r-process nucleosynthesis in astrophysical environments
like in supernovae deep interior via the ${\nu}_e$ and $\overline{\nu}_e$ reaction
processes:
\begin{equation*}
 {\nu}_{e} + n \rightarrow p + e^-  : {\rm rate}\, S_{{\nu}_e n} ; \hskip
0.2truecm \overline{\nu}_{e} + p \rightarrow n + e^+ : {\rm rate}\,
S_{\overline{\nu}_e p}  \, .
\end{equation*}
If indeed $\overline{\nu}_e$s could be over-abundant than $\nu_e$s, then, from the above expression one concludes that the neutron production is expected to be higher than the proton production in the supernova inner cores,  the sort of astrophysical sites we are focusing on in this paper as the supposedly last stage preceding the formation of the RNCBH. Thence, the theoretically well-known and proven supernova spin-flip conversion ${\nu}_e \longrightarrow \overline{\nu}_e$ (Majorana type neutrinos, for instance) could be significantly stimulated due to gravity-induced effects inside supernovae cores so as to possibly afford for the over-abundance of neutrons required for the r-process to effectively happen in this spacetime.

In providing the following estimates of the neutron-to-proton ratio we follow the
important paper by Fuller and Qian \cite{1996NuPhA.606..167F} (see also \cite{APJL789-M.Shibata-2014}). At a radial coordinate $r$ in the RNCBH spacetime, the electron fraction is determined as
in Eq. (\ref{e-fraction}) by the local values of the luminosities and average energies of
the ${\nu}_e$ and $\overline{\nu}_e$. However, since these neutrino species have
differing production/emission radii (i.e., their neutrinospheres have different values
of the RNCBH radial coordinate), they should undergo very different gravitational redshift
effects. If we define the ${\nu}_e$ neutrinosphere to be at  $r^{\nu-sp}_{\nu_e}$ and the
$\overline{\nu}_e$ neutrinosphere to be at $r^{\nu-sp}_{\overline{\nu}_e} $, then the second term of Eq.
(\ref{e-fraction}) can be recast as
\begin{equation}
Y_e = \frac{1}{ 1 + R_{\frac{n}{p}} }, \hskip 1.0truecm R_{\frac{n}{p}} \equiv
R^0_{\frac{n}{p}} \cdot \Gamma,
\label{gravit_Ye}
\end{equation}
with
\begin{equation}
R^0_{\frac{n}{p}} \simeq \left[\frac{ L^{\nu-sp}_{\overline{\nu}_e } \, \langle
E^{\nu-sp}_{\overline{\nu}_e} \rangle}{ L^{\nu-sp}_{\nu_e} \, \langle
E^{\nu-sp}_{\nu_e}\rangle}
\right]
\label{neurt-prot-ratio}
\end{equation}
In these equations, $ L^{\nu-sp}_{\overline{\nu}_e}, \langle E^{\nu-sp}_{\overline{\nu}_e}
\rangle$ are the average $\overline{\nu}_e$ energy and luminosity as measured by a
locally inertial observer at rest at the $\overline{\nu}_e$ neutrinosphere; and similarly
for the quantities which characterize the ${\nu}_e$ energy and luminosity at the ${\nu}_e$
neutrinosphere. (We recall that first order corrections in $a/r$ for this  spacetime do not affect local energy measurements.) The approximation is made so that the ${\overline{\nu}_e}, {\nu}_e$ energy
spectrum does not evolve significantly with increasing radius above the
${\overline{\nu}_e}, {\nu}_e$ sphere, as a result of the concomitant  emission, absorption
and scattering processes. The quantity $R^0_{\frac{n}{p}}$ is the local neutron-to-proton
ratio. All the above  quantities are understood to be evaluated from the neutrino and
antineutrino energy spectra extant at the RNCBH radial coordinate $r$.

In the above discussion the effects of the RNCBH gravitational field would be contained in the
parameter $\Gamma$, in the form

\begin{equation}
\Gamma \equiv \left[\frac{g_{00}(r^{\nu-sp}_{\overline{\nu}_e})}{
g_{00}(r^{\nu-sp}_{{\nu}_e})}\right]^{\frac{3}{2}}\label{Gamma}.
\end{equation}

Since in general as is concerned to the $u$ coordinate, the metric of a charged spacetime is larger than the Schwarzschild one for any $u<1/2$, and the ${\overline{\nu}_e}$ neutrinosphere is bigger than the ${\nu}_e$ neutrinosphere, from Eqs. (\ref{gravit_Ye}), (\ref{neurt-prot-ratio}) and (\ref{Gamma}) it follows that $Y_{e}^{Q}<Y_{e}^{Schw}$. This means that the $n/p$
associated with charged spacetimes are in general larger than their neutral counterparts for the supernova ejecta. This is a feature that naturally favors r-processes. Therefore, only the presence of charge per se may already
considerably change the neutron-to-proton ratio w.r.t. the Schwarzschild case. This should also be so for the case when nonlinear theories of the electromagnetism are compared to neutral solutions. This all means that in principle, due to the large number of neutrinos in supernova events, r-processes could reveal charged phases of black holes. The issue of assessing which theory is the one underlying a possible charged black hole seems also possible at first, due both to the aforementioned large number of neutrinos involved and the fact that supernova events are usually related to strong gravitational and electromagnetic fields, where spacetime metrics related to nonlinear Lagrangians should differ more significantly from their classical counterparts. In Sec. \ref{summary} we qualitatively discuss a way to probe that by means of light polarization measurements, as well as possible difficulties involved.

In the next section we perform some simple estimates on the r-process with the intent to evidence the importance of charged (and nonlinear) environments, in the context of the neutron-to-proton ratio in supernova ejecta and also as related to the neutrino oscillation lengths.

\section{Some simple estimates}
\label{estimation}
In our previous calculations we assumed that $a/r_{+}\ll1$, which is equivalent to considering $a/r\ll1$ or $a/M\ll 1$. In order to give an astrophysical application to this approximation, first consider the following system: a neutron star of mass $M$ and radius $R$, whose charged nucleus is ongoing a gravitational collapse with an oppositely charged crust that is left behind. The physical reasons that may lead to this scenario, as well as others resulting in transiently charged black holes, will be discussed in the next section. Suppose besides that such a charged
core spins rigidly with constant angular velocity, whose norm we take as $\Omega_{rot}$. If the system rotates slowly, in first order of
approximation we could take it as spherically symmetric. Therefore,
its angular momentum could be estimated as being proportional to $MR^2\Omega_{rot}$. It is simple to see that in taking into account general relativistic requirements, the constrain $a/r \ll 1$ can be cast as
\begin{equation}
\Omega_{rot} R\ll c\label{approxcond},
\end{equation}
which is naturally the same as in Ref. \cite{1967ApJ...150.1005H}. If we now take the radius of the stars to be of order of the Schwarzschild horizon, $R\approx 2MG/c^2$, then, from the above equation it follows that $\Omega_{rot}\ll c^3/(2MG)=10^{5}(M_{\bigodot}/M)$~Hz.
Actually, the fastest pulsar ever measured so far has rotation frequency around $720$ Hz (\cite{2014NuPhA.921...33B} and references therein).
Hence, our slow rotation description would be of relevance for several neutron stars. For an ordinary stable neutron star, with $M\approx M_{\bigodot}$ and $R_{\star}\approx 10^{6}$ cm, its Schwarzschild horizon is located at  $R_{schw}\approx 10^{5}$ cm. Let us posit that during the dynamical collapse of the star core, which satisfies Eq. (\ref{approxcond}), its crust (or charged magnetosphere) has remained at $R_{\star}$ (see the next section for further details). Then the latest neutrinos emitted by the inner core could travel up to 10 $R_{Schw}$ before interacting with the crust. In this region nonlinear effects could play a role. Assume, just as an example, that for the charged core $\alpha=0.5$ and $bM=0.017$. Take for the radial coordinate the value $u=M/r=0.45$ for the neutrino emission. In this case $g_{00}^{BI}(0.45)\approx 0.135$ and $g_{00}^{Ma}(0.45)\approx 0.151$. In these  conditions Eq. (\ref{osclength}), when brought to usual cgs units, becomes 
\begin{equation}
L_{osc}(cm)=\frac{123}{\sqrt{g_{00}}}\frac{E/MeV}{(\Delta m/eV)^2}\label{osclengthuu}.
\end{equation}
Assuming $\Delta m^2 \approx 0.01\, {\rm eV^2} $ and $E\approx$ MeV, we thus have that $L_{osc}^{BI}
\approx 3.35\times10^{6}$ cm, while $L_{osc}^{Ma}\approx 3.16\times10^{6}$ cm. Thereby, $(L_{
osc}^{Ma}/L^{BI}_{osc})^2\approx 0.8$, as it can be checked in Fig. \ref{lengthratio}.
For this case, there is a change of around $10\%$ in the oscillation lengths concerning the
Born-Infeld and Maxwell Lagrangians. Notice that this example gives an oscillation length
of the same order of distance as that one where the charged crust lies. Therefore, the different
theories chosen could dramatically change the
fate of the charged crust left behind, as well as for the envelope surrounding such star. Since
the number of neutrinos emitted in a neutron star system is colossal, even small changes on
the neutrino oscillation lengths, accounted for by {nonlinear} Lagrangians,  could play an
important role
into the evolution of the aforementioned system.

Let us make some estimates concerning $Y_e$ (and consequently $r$-processes) for astrophysical systems. Assume again that the masses involved in our problem are on the order of the solar mass. We take $r^{\nu-sp}_{{\nu}_e}= 3.5 $ Km and $r^{\nu-sp}_{\overline{\nu}_e}= 0.9 \, r^{\nu-sp}_{{\nu}_e}$, just to assume a case
where the neutrino spheres are closer to the outer horizon of some astrophysical system ($r_{+}\approx 3$ Km). This choice leads to $u_{\nu_e}=0.42$ and $u_{\overline{\nu}_e}=0.467$. Besides, we take \cite{1996NuPhA.606..167F} $E^{\nu-sp}_{\overline{\nu}_e}= 25$ MeV and $E^{\nu-sp}_{{\nu}_e}= 10$ MeV. Figure \ref{Yelarger} depicts $Y_e$ for such a case.
\begin{figure}[!hbt]
\centering
\includegraphics[width=1.\hsize]{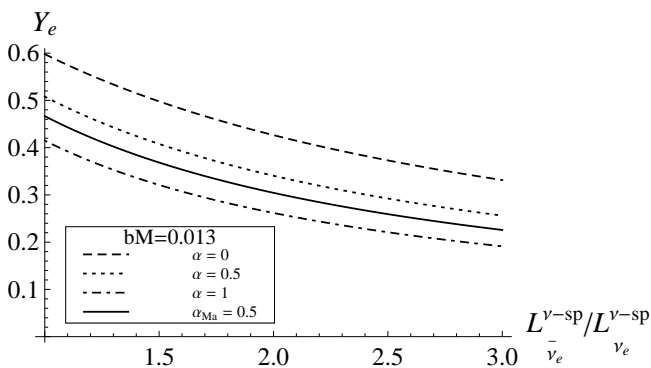}
\caption{{\small\sf$Y_e$ for the case $u=0.42$ as a function of the anti-neutrino-to-neutrino local luminosity. For this case, it was chosen $r^{\nu-sp}_{\nu_e}=3.5$ Km just to try to simulate the case where the neutrino sphere are close to the horizon of the collapsing system. One sees from this plot that $Y_e^{Schw}$ is larger than $Y_e^{BI}$ and $Y_e^{Ma}$. Further for a given $\alpha$, $Y_e^{BI}> Y_e^{Ma}$.}}
\label{Yelarger}
\end{figure}
In the scope of the Born-Infeld electromagnetism, it is not difficult to verify that $Y_{e}^{Sch}>Y_e^{BI}>Y_{e}^{Ma}$ (the latter two inequalities are naturally associated with a given $\alpha$). This means that charged black holes in the interior of the supernova envelopes would favor r-processes and therefore they could be a potential way to probe nonlinear electrodynamics.

\section{Discussion and Summary}
\label{summary}
We start by envisaging a method to disentangle rotation effects from charge ones within supernova observations, important in order to  advance with probes of nonlinear theories of the electromagnetism. It is well-known that distant supernovae appear only as point-sources of light, so asymmetric shapes could not be seen directly. Instead, they should be inferred from the way the light is polarized. In the light from a spherically symmetric star, however, all orientations are equally represented, and there is no net polarization. This is not the case for an asymmetric star or explosion. Light emitted along the longer axis shows a net excess of a particular polarization.

In 1982 Shapiro and Sutherland \cite{Shapiro1982}  introduced the concept of supernova asymmetry in astronomy. They purported that supernova (SN) atmospheres are scattering dominated,  based on the idea that light from an unresolved,  asymmetric, scattering atmosphere is linearly polarized.
In this seminal article they computed (modeled) the degree of linear polarization of light  from supernovae (SNe) which a nonspherical, scattering-dominated, supernova atmosphere had to produce as a function of its asphericity. It was claimed that any detection of net polarization of the supernova light should be a direct measure of the asphericity of its atmosphere, and that such a feature could affect what distant indicators, and several other astronomical parameters, could afford.
In our understanding, \cite{Shapiro1982} became a significant advance in the studies of stellar explosions, making of polarimetry a powerful tool in astrophysics, which has been  extensively in use so as to include astronomical radio observations \cite{Gaensler2015}.

Just let us quote one of those breakthrough observations where the polarization of light emitted from several supernovae have been measured. Wang and collaborators \cite{Wang2003} observed supernova 2001el which was brightened and dimmed. This was the first time the intrinsic polarization of a normal Type Ia supernova had been detected. This group was able to show that at peak brightness the exploding star was slightly flattened, with one axis shorter by about 10 percent. By a week later, however, the visible explosion was virtually spherical. Indeed, they claim that as spherical symmetry begins to dominate, about a week after maximum, it is not because the supernova is changing shape, but because we are seeing different layers of it. This way, outer layers expanding at thousands of kilometers per second would grow diffuse and become transparent, allowing the inner layers to become visible. They also stressed that when the star explodes, the outer part is aspherical, but as seeing lower down, the dense inner core appears spherical. 
 
Now, as concerns our study in the present paper, asphericity is the sort of geometrical effect one should expect from an astrophysical rotating compact object, or spacetime.  That is, if supernovae are not spherically symmetric, they should shine more brightly in one direction than in others. Thus, since neutrino oscillations can take place both at the supernova planar and radial directions, one could expect to have different contributions to the abundance of r-process products along the equatorial plane than at any other particular direction, for instance the polar direction. This would mean that the light from BH-forming type II supernovae and hypernovae would exhibit some degree of polarization due to rotation. Likewise, if a specialized method of densitometry could be performed in the observed supernova envelope, one could measure the r-products' abundance in each of such distinct spatial directions. For instance, for neutrino propagation in a slowly rotating spacetime along a circular orbit of radius $R$ in the equatorial plane, the local energy $E_l=E/\sqrt[]{g_{00}}$ ($E$ is a constant, the energy at infinity) is constant along this pathway. This is in contrast to the case of propagation in a radial trajectory. In these different directions the neutrino phases are given by [see the second equality of Eq. (\ref{osclength})]
\begin{equation}
\Phi_{\theta=\pi/2} =  \frac{m^2}{ E_l} R(\phi - \phi_0) \neq \Phi_{rad} =  \frac{m^2}{E_l} (r - r_0) \, ,
\label{S-T-Nu-Phases}
\end{equation}
where $\phi, \phi_0$ are angular positions, while $m$ is the neutrino mass. Clearly the phase in the second part of Eq.~(\ref{S-T-Nu-Phases}) is dominated by the gravitational redshift at positions $r, r_0$, whereas the planar phase is constant for a given $R$. This means that gravity has no effect in the latter, which is the same as for flat spacetime. Thenceforth, having available a  comprehensive sample of type II supernovae and hypernovae exhibiting light polarization, e.g. a supernova with noticeable asphericity and another with less or virtually spherical, might help discern on the role of rotation of the supernova progenitor and the just-formed BH in the magnitude of the enhancement of the r-process products.
 
On the other hand, the presence of an evolving electric field in the supernova ejecta could change the degree of polarization of light outcoming from the inner core.  As the strength of the electric field depends on the total charge which is generating it,  then comparing different degrees of polarization in samples of BH-forming type II supernovae and hypernovae may allow the disentanglement of their relative contributions to the total light polarization.
This way, implementing a detailed analysis of supernova samples exhibiting light polarization and the presence of electromagnetic fields (e.g. inferred from either Zeeman effect or Stark effect observations), one could have a tool for disentangling the role of rotation and charge (nonlinearities of electromagnetism)  from  astronomical observations. 
 
Indeed, due to neutralization aspects, one would expect that charged black holes are mainly related to unstable scenarios, which would thus lead them to be short-lived. In this regard, several physical mechanisms could be conceived for their formation. We envisage some here (for other mechanisms to generate black holes, not necessarily charged ones, see Ref. \cite{Woosley1996}).

Consider compact stars (neutron stars) that exhibit large magnetic fields and are good conductors (Goldreich-Julian's model \cite{Shapiro-book}). These systems are such that electric fields would also be present and would be of order $(\Omega R/c) B$ \citep{Shapiro-book}, where $\Omega$ is the angular frequency of the star, and $R$ and $B$ are its radius and magnetic field, respectively. From the above one immediately concludes that the larger the magnetic field the larger the electric field, whose origin would be related to a charge density that for certain cases would lead to a net charge. \footnote{For example, the discontinuity of the normal component of the electric field at the star's surface would be related to the charge density $-B_p\Omega R \cos^2\theta/(4\pi c)$ \cite{Shapiro-book}, $B_p$ the magnetic field at the pole of the star, which would result in a non-null net surface charge. Besides, when one neglects macroscopic currents near the surface of the star and assumes it has a permittivity close to that one of the vacuum, the electric field would be associated with the charge density $-\vec{\Omega}\cdot \vec{B}/(2\pi c)$ \cite{Shapiro-book}, which in general would also lead to a net charge.} For instance, a net charge of $10^{20}$ C ($\alpha \approx 0.1$ for stars with masses around a solar mass) would be related to a magnetic field of around $10^{19}$ G for (typical) neutron stars with $\Omega R/c \approx 10^{-2}$ and $R\approx 10^6$ cm. It is known that very large magnetic fields would lead compact stars to be unstable, {possibly collapsing into black holes \cite{Herman2003,Debora2014}. A neutron star with a mass around 2 solar masses and radius around 10 Km would be unstable for magnetic fields larger than $10^{18}$~G--this value is estimated by using the virial theorem in astrophysics \cite{Shapiro-book} and is related to fields of any nature (dipolar, poloidal or toroidal); stable systems are the ones in which their magnetic energies are smaller than the magnitude of their gravitational energies. Thus, for a star collapsing as a whole leaving behind a charged magnetosphere (this might happen due to their very different natures, which would imply very different characteristic times of collapse), the latter presumably always present in highly magnetized systems \cite{Shapiro-book}, short-lived charged black holes could always be formed. It is even possible that the crust may be left behind when the core collapses because it should interact more strongly with the magnetosphere. In all cases, the  typical sizes related to charged spacetimes would be of the order of the radius $R$ of the star and the times they would be charged are around $10^{-4}$ s ($\approx 1/\sqrt[]{G\rho}\approx R/c$ for core densities around $10^{15}$ g.cm$^{-3}$). We plan to investigate more carefully all these scenarios elsewhere.  

The above-mentioned mechanisms, as well as others, would motivate searches for (nonlinear) charged black holes and their natural ``probers'' would be neutrinos, given their bountiful production in any astrophysical context; see for instance Ref. \cite{Beacom-BH-SN}. As a by-product of this, in principle it would be possible to assess the nature of electromagnetism (Maxwell's or not), due to the special imprint different theories would have on certain phenomena, such as neutrino oscillations, spin-flip and r-processes in supernova events, as we have analyzed in this work and commented previously. One should bear in mind, though, that ambiguities may still arise regarding  probes of nonlinear theories of the electromagnetism, since they are intrinsically associated with (yet unknown) scale parameters as well as the charges and angular momenta of the transient black holes, neutrinospheres, neutrino luminosities, neutrino energies, neutrino masses, etc., which could all lead to overlaps in physical observables. Nonetheless, even in spite of these difficulties, it is worthwhile investigating at least consequences of charged black holes, because even if they are fleeting they could be astrophysically relevant.

In summary, we first solved generically Einstein's equations for slowly rotating black holes minimally coupled to nonlinear Lagrangians of the electromagnetism dependent upon its two local invariants. 
We used neutrinos (in the WKB approximation) to probe some of the aspects of these spacetimes, which may be invaluable tools to discern charged and uncharged black holes, as well as Maxwellian from nonlinear electrodynamics. The major departures from the classical case concerning the magnetic fields, the off-diagonal metric term, the precession of gyroscopes, the spin-flip, the neutrino flavor oscillation, etc. would just occur near the outer horizon of a nonlinear slowly rotating black hole because it would be there that the spacetime properties would change more pronouncedly. Besides, kinematical effects such as precessions (to be measured with gyroscope-like systems) could be of relevance in order to distinguish nonlinearities present in charged black holes, as well as experiments that take into account magnetic fields (asymptotically dipolar ones). Our calculations suggest that magnetic fields from nonlinear electrodynamics should deviate more pronouncedly apropos of their Maxwellian counterparts. Therefore, subsequent investigations on the probe and nature of charged black holes should focus more closely on this aspect.

Concerning the relevance of our analyses to supernova events, we have pointed out that the presence of charge only per se may considerably change the neutron-to-proton ratio in supernova ejecta apropos of neutral solutions to general relativity, which would already change the r-processes. This would mean that in principle nonlinear charged black holes could indeed influence more supernova events and the formation of heavier elements than Schwarzschild ones, which deserves better studies that we let to be elaborated elsewhere.

\acknowledgments
We thank the anonymous referee for valuable suggestions that helped us improve this manuscript. J.P.P. acknowledges the financial support given by Funda\c c\~ao de Amparo \`a Pesquisa do Estado de S\~ao Paulo (FAPESP) under the project No. 2015/04174-9. H.J.M.C. acknowledges the ICRANet Coordinating Headquarters in Pescara, Italy, for hospitality, and the support by the International Cooperation Program CAPES-ICRANet, financed by CAPES - Brazilian Federal Agency for Support and Evaluation of Graduate Education within the Ministry of Education of Brazil. HJMC also thanks the COLCIENCIAS PROGRAM ``Es Tiempo de Volver'' for financial support under the contract 390-2015 during part of the time required to produce this paper. G. Lambiase thanks the COST project CA15117 CANTATA.

\end{document}